\documentclass[]{JHEP3}
\usepackage{amsmath}
\usepackage{amssymb}
\usepackage{amsthm}
\usepackage{psfrag}
\usepackage{graphicx}

\newcommand{\be}{\begin{equation}}
\newcommand{\ee}{\end{equation}}
\newcommand{\beqa}{\begin{eqnarray}}
\newcommand{\eeqa}{\end{eqnarray}}

\newcommand{\KK}{{\cal K}}
\newcommand{\pd}{\partial}
\newcommand\m{\mu}

\newcommand\n{\nu}

\newcommand\s{\sigma}

\def\d{\partial}
\def\dpa{\partial^{\parallel}}
\def\dpe{\partial^{\perp}}
\def\Bpe{\Box^{\perp}}
\newcommand{\bseq}{\begin{subequations}}
\newcommand{\eseq}{\end{subequations}}

\renewcommand{\Im}{\mathop{\rm Im}\nolimits}

\newcommand{\di}{\mathrm d}

\title{Models of non-relativistic quantum gravity:\\
the good, the bad and the healthy}

\author{D. Blas,\!$^a$ O. Pujol\`as,\!$^b$ S. Sibiryakov,\!$^{a,c}$\vspace{.2cm}\\
\normalsize\llap{$^a$}
 \it FSB/ITP/LPPC,
 \'Ecole Polytechnique F\'ed\'erale de Lausanne,\\
 \normalsize\it CH-1015, Lausanne, Switzerland\\
 \normalsize\llap{$^b$}\it CERN, Theory Division, CH-1211 Geneva 23, 
Switzerland\\
\normalsize\llap{$^c$} \it Institute for Nuclear Research of the
Russian 
Academy of Sciences, \\
      \normalsize \it  60th October Anniversary Prospect, 7a, 
117312 Moscow, Russia}
\abstract{
Ho\v rava's proposal for non-relativistic quantum gravity introduces a preferred
time foliation of space-time which violates the local Lorentz
invariance. 
The foliation is encoded in a dynamical
scalar field which we call `khronon'.
The dynamics of the khronon field is sensitive to the symmetries and
other details of the particular implementations of the proposal.
In this paper we examine several consistency issues present in three
non-relativistic gravity theories:
Ho\v rava's projectable theory, the healthy non-projectable extension,
and a new extension related to ghost condensation. We find that 
the only model which
is free from instabilities and strong coupling is the non-projectable
one. We elaborate on the phenomenology of the latter model including a
discussion of the couplings of the khronon to matter. 
In particular, we obtain the parameters of the
post-Newtonian expansion 
in this model and show that they are compatible with current
observations.}

\begin{document}

%\newpage
%\tableofcontents
%\newpage

%%%%%%%%%%%%%%%%%%%%%%%%%%%%%%%%
\section{Introduction}
%%%%%%%%%%%%%%%%%%%%%%%%%%%%%%%%

Recently P. Ho\v rava argued that it may be
possible to construct a consistent renormalizable theory 
of quantum gravity within the
framework of perturbative quantum field theory (QFT)
\cite{Horava:2009uw} 
(see also
\cite{Horava:2008ih}). 
The proposal exploits the improved ultraviolet (UV) behavior of non-relativistic
QFTs possessing an UV fixed point with anisotropic scaling of space and
time. In this type of theories the UV behavior of the field
propagators is improved thanks to terms with higher spatial
derivatives. At the same time the number of time derivatives in the
Lagrangian remains equal to two allowing to bypass the problems with
ghosts arising in Lorentz invariant higher-derivative
theories of gravity \cite{Stelle:1977ry}. 
Evidently, by breaking the symmetry between space and time one
sacrifices Lorentz invariance (LI). The latter is 
no longer a
fundamental property of the theory and may only emerge at low energies
as
an approximate symmetry.
For matter fields in
flat space this does not pose an immediate problem,  
other than the stringent observational constraint that
Lorentz invariance needs to emerge to an extremely high
accuracy, see
\cite{Colladay:1996iz,Coleman:1998ti,Mattingly:2005re,Jacobson:2005bg,
Liberati:2009pf,Kostelecky:2008ts}
and references therein for the discussion of experimental bounds on  
Lorentz
violating extensions of the Standard Model.

However, 
abandoning relativistic invariance has 
 dramatic effects for gravity. The reason is that in general
relativity (GR) Lorentz symmetry is part of the local gauge group
whose role is to remove the unphysical degrees of freedom 
contained in the metric, leaving 
only two massless helicity-2 modes for the graviton. Thus one
expects that abandoning LI 
will lead to the appearance of new degrees of freedom in the
theory. As already pointed out in \cite{Horava:2009uw} this is indeed
the case: in addition to the helicity-2 modes the Ho\v rava model
propagates a helicity-0 excitation\footnote{The helicity-0 mode is
  absent in a special case when the model obeys anisotropic Weyl
  invariance. However, this symmetry is not compatible with
  observations. For another proposal to eliminate the
  helicity-0 mode see \cite{Horava:2010zj}.}. 
The physical meaning of the new mode can be understood as
follows. From the geometrical point of view the introduction of the
preferred time coordinate amounts to equipping the space-time manifold
with a foliation by space-like surfaces. In the gravitational theory the
foliation inevitably becomes dynamical together with the geometry of
the manifold. The helicity-0 mode is nothing but the excitation of
this foliation structure. In this sense, the extra mode describes
fluctuations of the global time, so we coin for it the name
`khronon'\footnote{From Greek $\chi\rho o\nu o\varsigma$ --
  time. This should not be confused with the term `chronon' appearing in
  the sense of ``fundamental interval of time'' in some theories 
\cite{Levi}.}.

Importantly, the new mode 
does not have a mass gap and thus cannot be consistently
decoupled at low energies \cite{Sotiriou:2009bx,bps1}. This implies 
that, contrary to the original expectation \cite{Horava:2009uw},
the theory cannot flow to GR at low energies. Instead, one may
entertain the possibility that at low energies the theory reduces to a
(Lorentz-violating) model of modified gravity, with the modifications
being small enough not to contradict the experimental data. 
The studies of modified gravity models, both in Lorentz-invariant 
\cite{Deffayet:2001uk,ArkaniHamed:2002sp,Luty:2003vm,Nicolis:2004qq} and 
Lorentz-violating
\cite{ArkaniHamed:2003uy,Dubovsky:2004sg,Blas:2009my} 
contexts, have shown that the
properties of the extra degrees of freedom can make
them fail as phenomenologically acceptable  effective field theories
(EFT). This 
is precisely what happens in the original realization
\cite{Horava:2009uw} of Ho\v rava's proposal where the behavior of the
khronon turns out to be pathological
\cite{Charmousis:2009tc,Li:2009bg,bps1,Koyama:2009hc,Henneaux:2009zb}.   

It is worth clarifying whether 
these pathologies are completely generic and invalidate the approach to quantum
gravity 
proposed in \cite{Horava:2009uw}, 
or they represent merely a failure of a specific
realization of the general framework. The purpose of the present paper
is to address this 
question. 

To this end we consider three different models of non-relativistic
gravity.  First we will reexamine the `projectable' version of the
original proposal \cite{Horava:2009uw} outlining its problems
associated with the scalar sector.  Then we will consider the two
extensions of the original proposal suggested in \cite{bps1} to
remedy these problems.  The first one is based on a smaller
symmetry group and can be viewed as a (power-counting) renormalizable
version of ghost condensation \cite{ArkaniHamed:2003uy}.  
Such a model
allows to alleviate the problems of the low-energy EFT.
However, we find that this leads to other
troubles.  The reason is the unavoidable presence in this
formulation of a second helicity-0 mode at high energies that 
leads either to fast instabilities or to the break-down of the perturbative
description. 
Hence, we conclude that this attempt does not offer
a promising candidate theory of quantum gravity.

The second extension is the one presented in \cite{bps2}.  In contrast
with the two previous models, in this case the scalar sector is free
from pathologies.
A preliminary study of this healthy extension was reported in
\cite{bps2,bps3}. Here we analyze it in more detail and confirm that
the model is compatible with phenomenological constraints for suitable
choices of parameters. This suggests that it may serve as a starting
point for constructing a renormalizable theory of quantum
gravity.

All the models we will consider contain flat space-time as a consistent 
background and are naively power-counting
renormalizable. However, the behavior of the extra mode(s) is
drastically different in the three cases, the differences stemming
from the symmetries or other details of the particular realization.
Hence, before going to our analysis let us spend a few words
concerning the main features of the different possible
implementations of non-relativistic quantum
gravity. One can distinguish the following: 
\begin{itemize}
\item The choice of the anisotropic scaling to implement
  power-counting renormalizability. Namely, one postulates the scaling
  transformations 
\be
\label{scal}
{\bf x}\mapsto b^{-1} {\bf x}~,~~~ t\mapsto b^{-z} t\;,
\ee
with a given critical exponent $z$, together with the scaling
weights of the different fields. Then one classifies the operators
in the theory according to their dimensions with respect to this
scaling. The theory is (power-counting) renormalizable if the action
contains only a finite number of terms of zero (marginal operators) and negative
(relevant operators) scaling dimensions\footnote{The dimension of the
  spatial coordinates $x_i$ is taken to be $-1$.}. The case of
a relativistic QFT corresponds to $z=1$. In this paper we stick to the
choice $z=3$. As discussed in \cite{Horava:2009uw}, this is the
minimal value of $z$ that allows to construct a power-counting 
renormalizable Lagrangian for gravity in $(3+1)$ dimensions.  Larger values
of $z$ are also possible and lead to super-renormalizable models.

\item The subgroup of the four-dimensional 
 diffeomorphisms (Diff) under which the theory is
  invariant.  The distinction between space and time
   enforces a preferred frame, and so a preferred time
  coordinate. As stated above, this corresponds to
  endowing the space-time manifold with an additional structure:
  a preferred foliation by space-like surfaces.
 In particular this means that the
  arbitrary reparameterizations of time $t \mapsto \tilde t(t,{\bf
    x})$ are not an invariance of the theory. Instead, 
the following unbroken symmetries have been considered in the
literature: 
\begin{enumerate}
\item[(i)]~
\vspace{-19pt}
\be
\label{FDiffs}
{\bf x}\mapsto \tilde {\bf x} (t,{\bf x})~~\text{and}~~t \mapsto \tilde
  t(t).~~~~~~~~~~~~~~~~~~~~~~~~~~~~~~~~~~~~~~~~~~~~~~~~~~
\ee
We will refer to these transformations as 
`foliation-preserving Diffs', or FDiffs for short. 
This is the largest possible unbroken gauge group. 
It is the one originally considered in \cite{Horava:2009uw}.
\item[(ii)]~
\vspace{-19pt}
\be 
\label{RFDiffs}
{\bf x} \mapsto \tilde {\bf x} (t,{\bf x})
~~\text{and}~~t \mapsto\tilde t=t+const.
~~~~~~~~~~~~~~~~~~~~~~~~~~~~~~~~~~~~~~
\ee
We shall refer to this as the `restricted-foliation-preserving Diffs'
  (RFDiffs).  This symmetry arises in a number of effective field
  theories of modified gravity, such as 
 the shift-symmetric $k-$essence \cite{ArmendarizPicon:1999rj} or the
  ghost condensation \cite{ArkaniHamed:2003uy}.  The invariance under time
  translations implies existence of a conserved energy.
\item[(iii)]~
%\vspace{-20pt}
  ${\bf x} \mapsto \tilde {\bf x} (t,{\bf x}).$
  
  This is the unbroken group in potential-driven inflation and in
  non-shift-sym\-met\-ric $k-$inflation around time-dependent spatially
  homogeneous 
  solutions. It serves as the basis of the effective field theory 
 of inflation
  \cite{Cheung:2007st,Weinberg:2008hq}. The action in this case
  contains explicit time dependence and there is no energy
  conservation.
\item[(iv)]~
%\vspace{-20pt}
  ${\bf x} \mapsto \tilde {\bf x} ({\bf x})~~\text{and}~~
t \mapsto\tilde t=t+const.$

  This option is
  realized in Einstein-aether theory \cite{Jacobson:2000xp} (see
  \cite{Jacobson:2008aj} for recent review) and gauged
  ghost condensation \cite{Cheng:2006us}. It leads to the appearance of
  propagating helicity-1 degrees of freedom (in addition to the
  helicity-2 and helicity-0 modes).
\item[(v)]~
%\vspace{-20pt}
${\bf x}\mapsto \tilde{\bf x}={\bf x}+\mathbf{\xi}(t)
~~\text{and}~~t \mapsto\tilde t=t+const.$

Here $\xi(t)$ is and arbitrary time-dependent three-vector. This is the
symmetry group of Lorentz violating massive gravity
\cite{Dubovsky:2004sg,Rubakov:2008nh}. 
\end{enumerate}

In principle it may be possible to construct
power-counting renormalizable theories of gravity with any of the
above unbroken symmetries. A natural expectation is that 
the larger the unbroken
gauge group, the more constrained the model is, and the fewer degrees
of freedom it contains. 
Investigating all the possibilities is beyond
the scope of this article. 
We shall limit the analysis 
to the
FDiff-invariant and RFDiff-invariant
theories. 
In these cases there are only extra scalar
modes, but their number grows when relaxing the symmetry. We will see
that the covariant form of these theories involves only a single
scalar field (a khronon)
in addition to the metric.\footnote{The equation of motion for the
  khronon field turns out to be of different order in time-derivatives
depending on the symmetry. Hence the different number of propagating
modes in the FDiff- and RFDiff-invariant cases.} It seems therefore
appropriate to refer to these as ``khrono-metric'' theories. 

\item Finally, one may impose additional restrictions on the action
  that do not follow from a symmetry. Examples of such restrictions
  are  the `projectability' and the `detailed balance'
  conditions of the original paper \cite{Horava:2009uw}. In the
  present work we do not impose the detailed balance condition, 
  and 
  we
  consider models both with and without the projectability property.
\end{itemize}

The paper is organized as follows. In Sec.~\ref{sec:setup} 
we introduce the basic notations and tools for our analysis.
In Sec.~\ref{sec:proj} we consider the `projectable'
version of the original Ho\v rava model and discuss problems
associated with the extra scalar mode. In Sec.~\ref{sec:ghc} we
elaborate on an attempt to fix these problems by relaxing the
symmetry from FDiffs down to RFDiffs. We 
show that the model obtained in this way does not provide a
consistent candidate for quantum gravity.
The relation between this model and the
ghost condensation
\cite{ArkaniHamed:2003uy} is discussed.
In Sec.~\ref{sec:healthy} we return to the case of full
FDiff-invariance, now without the projectability condition.
In this case the original action of Ho\v rava's model must be
supplemented by additional terms \cite{bps2} allowed by the symmetry
and power-counting renormalizability. 
In
Sec.~\ref{sec:healthy1} we demonstrate that these terms
make the scalar graviton stable and weakly coupled. In the rest of
Sec.~\ref{sec:healthy} we analyze the phenomenological bounds on the
healthy model. In doing this we exploit the analogy
\cite{bps3,Jacobson:2010mx} 
between the low-energy limit of the model and the Einstein-aether
theory \cite{Jacobson:2000xp,Jacobson:2008aj}. 
In Sec.~\ref{sec:concl} we summarize our results and discuss future directions.
Some details of the analysis are deferred to the
Appendices. 

Readers interested in the phenomenology of the
healthy model may skip Secs.~\ref{sec:proj}, \ref{sec:ghc} and go
directly from Sec.~\ref{sec:setup} to Sec.~\ref{sec:healthy}.

%\newpage

%%%%%%%%%%%%%%%%%%%%%%%%%%%%%%%%
\section{General Setup}
\label{sec:setup}
%%%%%%%%%%%%%%%%%%%%%%%%%%%%%%%%

\subsection{Three theories under scrutiny}
\label{sec:theories}

In this section we introduce the basic ingredients of the Ho\v
rava-type theories that we are going to consider. The field content
includes
the
spatial metric $\gamma_{ij}$, the shift $N_i$ and the lapse $N$
entering into the (3+1) (ADM) decomposition of the
4-dimensional metric,
\[
\di s^2=(N^2-N_i N^i) \di t^2-2N_i  \di x^i \di t-\gamma_{ij}\di x^i \di x^j\;~.
\]
These fields transform in the standard way under the 4-dimensional
coordinate transformations.

%\newpage
\subsubsection*{Model I: Ho\v rava's projectable FDiff  gravity.}

We first consider the FDiff-invariant case.  Let us also impose the additional
requirement that the lapse is `projectable', i.e. that it does not
depend on space coordinates, 
\be
\label{pro}
N=N(t)\;.
\ee 
Note that
this restriction is compatible with the transformation rules for the
lapse under FDiffs,
\[
%\label{lapsetr}
N\mapsto\tilde N=N\,\frac{\d t}{\d\tilde t}\;.
\]
One writes down the following action with
two time derivatives
%\footnote{One
%could in principle consider actions of the form
%\be
%\label{toomuch}
%S_I=\frac{M_P^2}{2}\int \di ^3x\, \di t \sqrt{\gamma}\,N\,\big(K^{kl}P[\nabla_s,\gamma^{mn}]_{klij}K^{ij} - {\cal V}_I \big)\;.
%\ee
%with $P[\nabla_s,\gamma^{mn}]_{klij}$ being  a local operator depending on
%$\nabla_s$ and $\gamma^{mn}$. However, one can prove that
%if under the scaling (\ref{scal}) the operators of the 
%form $K\Delta^\alpha K$ (for $\alpha\geq 1$) 
%are marginal, the operators of the type 
%\oo{
%${\dot K}^2$
%} 
%will be also marginal or relevant. This means
%that, as happens in the Diff invariant case \cite{Stelle:1977ry}, 
%the renormalizable theory will include terms with four time derivatives, 
%which implies the appearance of ghosts in the spectrum. 
%We thank M. Asorey for discussions on this possibility.}
\cite{Horava:2009uw},
\be
\label{ADMact1}
S_I=\frac{M_P^2}{2}\int \di ^3x\, \di t \sqrt{\gamma}\,N\,\big(K_{ij}K^{ij}-
\lambda K^2 - {\cal V}_I \big)\;.
\ee
Here $M_P$ is the Planck mass; $\lambda$ is a dimensionless constant;
$K_{ij}$ is the extrinsic curvature tensor
for the surfaces of constant time,\footnote{Throughout the paper we
  use lower-case Latin letters $i,j,\ldots$ for 3-dimensional
  indices. They are raised and lowered using the spatial metrics
  $\gamma_{ij}$, $\gamma^{ij}$. The covariant derivatives carrying
  these indices are understood accordingly.} 
\be
\label{extr}
K_{ij}=\frac{1}{2N}\left(\dot\gamma_{ij}-\nabla_i N_j-\nabla_j N_i\right)\;,
\ee
$K$ is its trace, $K\equiv K_{ij}\gamma^{ij}$; and the `potential'
${\cal V}_I$ depends on the spatial metric $\gamma_{ij}$ via the
3-dimensional Ricci tensor $R_{ij}$ and its spatial covariant
derivatives. One notices that the first two terms in (\ref{ADMact1}),
which comprise the kinetic part of the action, are invariant under the
scaling (\ref{scal}) with $z=3$ provided $\gamma_{ij}$, $N_i$ and $N$ scale
as
\[
%\label{scalfields}
\gamma_{ij}\mapsto\gamma_{ij}~,~~~N_i\mapsto b^2 N_i~,~~~
N\mapsto N\;.
\]
In other words, the kinetic terms are marginal with respect to the
anisotropic scaling with $z=3$. 
A power-counting renormalizable theory
is obtained by including in the action all possible marginal and
relevant terms. This corresponds to considering the most general  
potential ${\cal V}_I$ containing local
operators with the scaling dimensions up to 6. There is 
a finite number of these terms, whose classification was performed in
\cite{Sotiriou:2009bx}. For our purposes it suffices to write 
schematically,\footnote{Recall that
  in three dimensions the Riemann tensor is completely
 determined in terms of the Ricci tensor.}
\be
\begin{split}
\label{potential}
{\cal V}_{I}=-\xi R &+M_*^{-2}(A_1 R^2+A_2R_{ij}R^{ij}+\ldots)\\
&+M_*^{-4}(B_1 R\Delta R + B_2 R_{ij}R^{jk}R_k^i+\ldots)\;,
\end{split}
\ee
where $\xi$, $A_n$, $B_n$ are dimensionless coupling constants and
dots 
stand for all the possible inequivalent operators of the given dimension.
Note that we have introduced here the mass  $M_*$
for the scale suppressing higher-order operators; this scale may or
may not coincide with $M_P$. 
In what follows we set the parameter $\xi$ to 1 which can always be
achieved by a constant rescaling of the time coordinate. In the
absence of matter this does not affect the physical content of the
model. We will return to the general case $\xi\neq 1$ in
Sec.~\ref{sec:healthy3}. 

Omitting higher-derivative terms in the 
potential and setting
$\lambda=1$ one formally recovers the action of 
GR \cite{Horava:2009uw}.\footnote{Still, the resulting theory is different
from GR because of the projectability restriction (\ref{pro}).}
However, as we discuss below the limit $\lambda\to 1$, $A_n, B_n\to 0$
is discontinuous: instead of flowing to GR the model (\ref{ADMact1}),
(\ref{potential}) becomes pathological in this limit. 
This model will be analyzed in
Sec.~\ref{sec:proj}.

\subsubsection*{Model II: non-projectable FDiff gravity}

Let us now relax the projectability condition (\ref{pro}), 
so that the lapse $N$
is allowed to depend both on time and space. In this case, the object  
\[
\label{ai}
a_i\equiv N^{-1}\d_iN\;,
\]
transforms covariantly under FDiffs and has  scaling dimension 
1. To obtain a (power-counting) renormalizable theory one should
allow the potential to depend on $a_i$. Again, this dependence must be
limited to local operators of dimension up to 6. Note that $a_i$ can be
consistently excluded from
the kinetic part of the Lagrangian. Indeed, recall that each time
derivative raises the dimension of an operator by 3. Thus operators of
dimensions less or equal 6 can contain together with $a_i$ at most one
time derivative. It is straightforward to check that, up to
integration by parts in the action, there are three possible
combinations (all having dimension 5):
\[
K^{ij}a_ia_j~,~~~K^{ij}\nabla_ia_j~,~~~K\nabla_i a^i\;.
\]
These operators are odd under T and CPT transformations. While in
general it might be interesting to study the effect of such operators,
in the present paper we forbid them by
assuming the CPT invariance.
Thus we arrive to the
following action
\be
\label{ADMact2}
S_{II}=\frac{M_P^2}{2}\int \di ^3x\, \di t \sqrt{\gamma}\,N\,\big(K_{ij}K^{ij}-
\lambda K^2 - {\cal V}_{II} \big)\;,
\ee
where 
\begin{align}
\label{potadd}
{\cal V}_{II}={\cal V}_{I}-\alpha\, a_ia^i
&+M_*^{-2}(C_1a_i\Delta a^i+
C_2(a_ia^i)^2+C_3 a_ia_jR^{ij}+\ldots)\\
&+M_*^{-4}(D_1a_i\Delta^2 a^i+
D_2(a_ia^i)^3+D_3 a_ia^ia_ja_kR^{jk}+\ldots)\;.\notag
\end{align}
Note that due to the positive scaling dimension of $a_i$ the potential
again contains only finite number of terms. The projectable model can
be recovered from (\ref{potadd}) by taking the limit $\alpha\to\infty$
which enforces the spatial gradient of the lapse to vanish, $\d_i
N=0$. The potential (\ref{potadd}) was first proposed in \cite{bps2}. 
 The model with the action (\ref{ADMact2}),
(\ref{potadd}) will be studied in Sec.~\ref{sec:healthy}.

\subsubsection*{Model III: RFDiff  gravity}

The third model we are going to consider corresponds to relaxing the
symmetry from FDiffs to RFDiffs. This has a dramatic effect on
the possible structure of interactions in the theory. The reason is that
the lapse $N$ is now a scalar under the symmetry group\footnote{Note
  that
the reduced gauge invariance is compatible with restricting
the lapse $N$
  to a fixed value, say $N=1$. Then $N$ drops out of the action. 
The only difference of the resulting theory from the projectable
model is the absence of the integral Hamiltonian constraint, so
locally the two theories are equivalent (cf. the discussion in the beginning of
Sec.~\ref{sec:projstr}). Thus all results about local dynamics of the
projectable model apply to this case.}. At 
the same time its scaling dimension is 0. Therefore, all dimensionless
couplings in
the Lagrangian may acquire arbitrary dependence on it without spoiling
the power-counting renormalizability. Besides, the reduced
symmetry allows and power-counting renormalizability requires
to include into the action a kinetic term for $N$. Thus 
the most general action reads,
\be
\label{ADMact3}
S_{III}=\frac{M_P^2}{2}\int \di ^3x\, \di t \sqrt{\gamma}\,N\,
\Big(\lambda_1(N)
(\dot N-N^i\pd_i N)^2
+\lambda_2(N)K_{ij}K^{ij}-
\lambda(N) K^2 - {\cal V}_{III} \Big)\;
\ee  
with
\begin{align}
\label{potIII}
{\cal V}_{III}=&V(N)-\xi(N)R-\alpha(N)\, a_ia^i\\
&+M_*^{-2}\big(A_1(N)R^2+A_2(N)R_{ij}R^{ij}+C_1(N)a_i\Delta a^i+
C_2(N)(a_ia^i)^2+\ldots\big)\notag\\
&+M_*^{-4}\big(B_1(N)R\Delta R+B_2(N)R_{ij}R^{jk}R_k^j+D_1(N)a_i\Delta^2 a^i+
D_2(N)(a_ia^i)^3+\ldots\big)\;.\notag
\end{align}
Note that the form of the kinetic term for $N$ is fixed by the
invariance under RFDiffs (\ref{RFDiffs}). Besides, relaxing the
symmetry down to RFDiffs allows to include the standard potential
terms $V(N)$ for the lapse. 
The presence of arbitrary functions of $N$ in the Lagrangian  makes 
this theory perhaps less attractive, since it involves 
an infinite number of coupling constants. 
In spite of this, we find it instructive to 
explore what happens in this type of extension.
We study this model in Sec.~\ref{sec:ghc}.
\\

Let us end this section by introducing a few notations that will be
common in the analysis of all the above models.  
We are going to study the dynamics of small metric perturbations in
these models 
above
flat background. Since the most worrisome modes are the
helicity-0 excitations we concentrate on scalar
perturbations of the metric which we parameterize as follows:
\bseq
\label{dec}
\begin{align}
\label{dec1}
&N=1+\phi\;,\\
\label{dec2}
&N_i=\frac{\d_i}{\sqrt{\Delta}}B\;,\\
\label{dec3}
&\gamma_{ij}=\delta_{ij}-2\left(\delta_{ij}-\frac{\d_i\d_j}{\Delta}\right)\psi\;
-2\frac{\d_i\d_j}{\Delta}E~,
\end{align}
\eseq
where $\Delta$ is the spatial Laplacian. 
Finally, the following mass scales will appear
in the analysis:
\be
\label{Mscales}
M^2_\lambda\equiv(\lambda-1) M_P^2~,~~~
M^2_\alpha\equiv \alpha M_P^2~,~~~
M^2_{\lambda_1}\equiv \lambda_1 M_P^2\;.
\ee

%%%%%%%%%%%%%%%%%%%%%%%%%%%%%%%%%%%%
\subsection{St\"uckelberg formalism and the khronon field}
\label{sec:stuck}
%%%%%%%%%%%%%%%%%%%%%%%%%%%%%%%%%%%%

In our study we will make extensive use of 
the St\"uckelberg formalism. This formalism
 allows to single out explicitly the extra degrees of freedom
 appearing due to 
breaking of Diff invariance and study their
dynamics in a transparent way. In the present context, it amounts
to rewriting the
action of the theory in a generally covariant form at the expense of
introducing a compensator field which transforms non-homogeneously
under the broken part of the 4-dimensional Diffs. It is worth
stressing that this procedure does not introduce new degrees of
freedom into the theory, but just makes explicit those already
present. See \cite{ArkaniHamed:2002sp,Dubovsky:2004sg,Rubakov:2008nh} 
for previous applications of the
St\"uckelberg formalism in modified gravity theories.

In the present paper we follow the approach of \cite{bps1}. One 
encodes the foliation structure in a
scalar field $\varphi(x)$. Namely, the foliation surfaces are identified
with the surfaces of constant $\varphi$,
\be
\label{fol}
\varphi=const\;.
\ee
Clearly, to define a regular foliation the field $\varphi$ must
possess non-zero time-like gradient. In this sense $\varphi$ defines
an absolute time, hence we call this field `khronon'.
This identification allows
a covariant definition of the foliation structure.
The actions
(\ref{ADMact1}), (\ref{ADMact2}), (\ref{ADMact3}) correspond to the
frame where the coordinate time coincides with $\varphi$,
\be
\label{unitg}
\varphi=t\;.
\ee
We will refer to this choice of coordinates as the `unitary gauge'.

The action for $\varphi$ in a generic frame is obtained by realizing that 
the objects appearing in (\ref{ADMact1}), (\ref{ADMact2}), (\ref{ADMact3})
are the standard geometrical quantities that characterize 
the embedding of the hypersurfaces (\ref{fol})
into space-time. The central role in the
construction of these quantities is played by the unit normal 
vector\footnote{The Greek indices
  $\mu,\nu,\ldots$ are raised and lowered using 
the 4-dimensional metric $g_{\mu\nu}$. 
The same correspondence applies to the covariant
derivatives carrying these indices.} 
\be
\label{umu}
u_\mu \equiv \frac{\partial_\mu\varphi}{\sqrt{X} }~.
\ee
where
\be
\label{X}
X\equiv g^{\mu\nu}\,\partial_\mu\varphi\,\partial_\nu\varphi~.
\ee
Note that $u_\mu$ is invariant under the reparameterizations of
$\varphi$,
\be
\label{phirepar}
\varphi\mapsto\tilde\varphi=f(\varphi)\;, 
\ee 
where $f$ is an
arbitrary monotonic function. This reflects the invariance of the
foliation structure under reparameterizations of $\varphi$.
The time-dependent VEV of the khronon field $\varphi$
breaks the product of
the reparameterizations
(\ref{phirepar}) and general covariance down to the diagonal
subgroup which in the unitary gauge coincides with the FDiffs
(\ref{FDiffs}). 
One concludes that the covariant form of a
FDiff-symmetric theory must be invariant under (\ref{phirepar}).
On the other hand,
in the case of RFDiffs the symmetry of the khronon action is reduced
to the shift symmetry
\be
\label{shift}
\varphi\mapsto\tilde\varphi=\varphi+const\;,
\ee
allowing
general dependence of the action on
the derivatives of $\varphi$.
 
Other geometrical quantities associated to
the foliation are constructed out of $u_\mu$ and its derivatives. We
have the following expressions for the spatial projector: 
\[
%\label{proj}
P_{\mu\nu}\equiv g_{\mu\nu}-u_\mu u_\nu\;,
\]
the extrinsic curvature:
\be
\label{extr1}
\KK_{\mu\nu}\equiv
P_{\rho\mu} \nabla^{\rho} u_{\nu}=
\frac{1}{\sqrt{X} } P_{\mu}^\rho P_{\nu}^{\sigma}\nabla_\rho
\nabla_\sigma \varphi\;,
\ee
the acceleration  of the congruence of curves normal to the foliation:
\be
\label{amu}
a_\mu\equiv u^\nu\nabla_\nu u_\mu
\ee
and
the intrinsic Riemann tensor:
\[
%\label{gauss}
{\cal R}^\mu_{~\nu\rho\sigma} = 
P^\mu_\alpha \,P_\nu^\beta\, P_\rho^\gamma\, P_\sigma^\delta\, 
{}^{(4)}R^\alpha_{~\beta\gamma\delta}
+ \KK^\mu_\rho \KK_{\nu\sigma} - \KK^\mu_\sigma \KK_{\nu\rho}~,
\]
where in the last equation ${}^{(4)}R^\alpha_{~\beta\gamma\delta}$ is
the 4-dimensional Riemann tensor.
Now it is straightforward to obtain
the covariant form of the actions (\ref{ADMact1}), (\ref{ADMact2}),
(\ref{ADMact3})  by 
identifying the quantities appearing in the ADM decomposition with 
the appropriate combinations of $\d_\mu\varphi$, $u_\mu$, $P_{\mu\nu}$,
$\KK_{\mu\nu}$, etc. Namely, writing the covariant objects in the
unitary gauge we obtain:
\begin{align}
&X=N^{-2}\;,\notag\\
%\label{uunit}
&u_0=N~,~~~u_i=0\;,\notag\\
%\label{Punit}
&P^{00}=P^{0i}=0~,~~~P^{ij}=-\gamma^{ij}\;,\notag\\
%\label{Kunit}
&\KK_{ij}=K_{ij}~,~~~\text{etc.}\notag
\end{align}
In this way one ends up with an action describing GR plus a derivatively
coupled khronon field $\varphi$. 

Postponing the detailed analysis to the following sections, let us discuss
here an important issue about the counting
 of degrees of freedom in the St\"uckelberg picture. 
Note that the higher \emph{spatial} derivatives of
Ho\v rava theories in the unitary gauge translate 
in the covariant picture into higher
\emph{covariant} derivatives of the khronon field. Consider, for example, the
covariant form of the action (\ref{ADMact2})
\be
\label{covar2}
S_{II}=\frac{M_P^2}{2}\int \di^4x\sqrt{-g}\Big\{-{}^{(4)}R
-(\lambda-1)\KK^2-\alpha\, a_\mu a^\mu
+\ldots\Big\}\;, 
\ee
where $\KK\equiv\KK_\mu^\mu$, and
$\KK_{\mu\nu}$ and $a_\mu$ depend on the khronon field via the
expressions (\ref{extr1}), (\ref{amu}), (\ref{umu}); dots stand
for the higher-order terms\footnote{We remind that we
  have set the constant $\xi$ in the potential (\ref{potential}) to
  $1$.}. It is easy to see that the second and the
third terms in (\ref{covar2}) contain four derivatives\footnote{The
  higher order terms omitted in (\ref{covar2})
 contribute with even more derivatives.} 
of $\varphi$. Thus in a general coordinate frame the equations of motion
for $\varphi$ are fourth-order in time. Naively, this implies presence
of ghosts and hence inconsistency of the theory. However, as discussed
in \cite{bps1} this conclusion would be incorrect due to the following
reason.  
Though the action (\ref{covar2}) has a covariant form, the physical
content of the model does depend on the choice of the coordinate
frame. The counting of degrees of freedom of the model must be
performed in the preferred frame which is set by the preferred
foliation. In this frame the number of time derivatives in the
equations following from (\ref{covar2}) is reduced to two, and the
ghost modes are absent.

While the above statement is obvious in the unitary gauge where
perturbations of the khronon field are set to zero, it is rather
non-trivial for general gauges where $\varphi$ is allowed to
fluctuate. Let us formulate this statement more precisely.
Consider perturbations of 
the field $\varphi$ around a background $\bar\varphi$,
\[
%\label{phipert}
\varphi=\bar\varphi+\chi\;.
\]
We do not make any assumptions about the background $\bar\varphi$, in
particular,
it
does not need to obey equations of motion. For the sake of the
argument
we treat the metric as external assuming that it remains unperturbed; 
again, the background value
$\bar g_{\mu\nu}(x)$ can be arbitrary.
Let us now choose a coordinate frame such that the time coordinate
coincides with the background value of the khronon field,
\be
\label{unitback}
t=\bar\varphi\;.
\ee 
Then in this frame
the linearized equations of motion for $\chi$ following from
(\ref{covar2}) are second order in time. 

This proposition is proved in
Appendix~\ref{app:A}. 
As explained there, the proof 
uses in an essential way the invariance of
the action under the transformations (\ref{phirepar}).
As a consequence it also applies to
the projectable model
(\ref{ADMact1}) which, being FDiff-invariant, obeys this symmetry. 
On the
other hand, the above statement does not hold for the model
(\ref{ADMact3}) whose covariant form is not invariant under
(\ref{phirepar}). 
We will see in
Sec.~\ref{sec:ghc} that this makes the equation of motion for
the khronon in the model (\ref{ADMact3})
fourth-order in time; as one can anticipate, this will lead to
certain pathologies.

%%%%%%%%%%%%%%%%%%%%%%%%%%%%%%%%%%%%%%
\section{Ho\v rava's projectable model} 
\label{sec:proj}
%%%%%%%%%%%%%%%%%%%%%%%%%%%%%%%%%%%%%%

\subsection{Gradient instability}

In this section we will study the projectable case
(\ref{ADMact1}), where the lapse $N$ is assumed to obey the
 condition (\ref{pro}).
Let us analyze the dynamics of 
small perturbations around flat background. At
this stage it is convenient to work in the unitary gauge. 
We concentrate on the scalar
perturbations of the metric and use the decomposition (\ref{dec}),
where
due to the projectability condition 
$\phi$ depends only on time.
Substituting these expression into the action (\ref{ADMact1}) we
obtain the following quadratic Lagrangian:
\be
\label{L2proj}
\begin{split}
{\cal L}^{(2)}_I=\frac{M_P^2}{2}\bigg\{&-2\dot\psi^2
-2 \psi\Delta\psi+4\psi\sqrt\Delta\dot B+4\psi\ddot E \\
&-(\lambda-1) \big(\sqrt\Delta B+\dot E+2\dot\psi\big)^2 
-\frac{f_1}{M_*^2}(\Delta\psi)^2
-\frac{g_1}{M_*^4}\psi \Delta^3\psi
\bigg\} \;,
\end{split}
\ee
where the constants $f_1$ and $g_1$ are related to the coefficients in
the potential (\ref{potential}); the precise form of this relation is
not important for us. Note that the perturbation of the lapse $\phi$
drops out of the quadratic Lagrangian because of the projectability
condition: when $\phi$ does not depend on the space coordinates all
terms containing it are total derivatives.

Integrating out the non-dynamical fields
$B$ and $E$, one finds 
\be
\label{L2proj_2}
{\cal L}^{(2)}_I=\frac{M_P^2}{2}
\left\{\frac{2(3\lambda-1)}{\lambda-1}\dot\psi^2
-\psi\left(2\Delta+\frac{f_1}{M_*^2}\Delta^2
+\frac{g_1}{M_*^4}\Delta^3\right)\psi\right\}\;.
\ee
Let us discuss this result. The positivity of the
kinetic term for the field $\psi$, needed for the positivity of kinetic
energy, 
imposes the restriction:
\be
\label{lambdarestr}
\frac{3\lambda-1}{\lambda-1}>0\;.
\ee
On the other hand, 
the dispersion relation following from (\ref{L2proj_2}) has the
form
\be
\label{disp}
\omega^2=-\frac{\lambda-1}{3\lambda-1}\left({\bf p}^2
-\frac{f_1}{2M_*^2}{\bf p}^4+\frac{g_1}{2M_*^4}{\bf p}^6\right)\;.
\ee
We see that for $\lambda$ satisfying (\ref{lambdarestr}) the field $\psi$ is
tachyonic at small values of spatial momenta. This instability can be cut off by
the second term in the bracket at 
$p\sim M_*$. One estimates the maximal rate of instability as
\be
\Im \omega \sim \sqrt{|\lambda-1|}\, M_*\;.
\ee 
To obtain an acceptable phenomenology,  we must require that the
instability does not develop during the life-time of the
Universe,\footnote{It was argued in 
\cite{Mukohyama:2009tp,Izumi:2009ry} that the instability may
  be cut off by non-linear effects which may lead to weakening of the
  bound (\ref{instbound}). However, this scenario looks
highly non-trivial and a detailed analysis of the
  non-linear dynamics of the model is needed to see if it is indeed
  realized. 
 Thus we prefer to stick to the bound (\ref{instbound})
  following from the linear theory.}
\be
\label{instbound}
\sqrt{|\lambda-1|}\,M_* < H_0\;,
\ee
where $H_0$ is the present Hubble expansion rate.
Recall that the physical meaning of $M_*$ is
that of the scale suppressing higher-derivative operators in the
gravitational action. In other words, it is the genuine scale of
quantum gravity in the model at hand. A theoretically motivated value
for $M_*$ would be a microscopic scale of order or somewhat below the
Planck mass. However, to be as general as possible, we choose not to
impose any a priori assumptions about the value of $M_*$. Then we are
left with the experimental lower bounds on $M_*$. 
The direct bound is rather mild and 
comes from the tests of Newton's law at the distances
$\sim 10 \mu$m \cite{Will:2005va} and reads,
\be
\label{M*bound}
M_*\gtrsim 0.1~\mathrm{eV}\;.
\ee
Even taking $M_*$ at the lower end of the allowed range, the stability
condition (\ref{instbound}) implies that the parameter $\lambda$ must
be extremely close to 1,
\be
\label{lprojbound}
|\lambda-1|\lesssim 10^{-61}\;.
\ee
This is unappealing from the theoretical viewpoint. In particular, it
is unclear if (\ref{lprojbound}) can be preserved by
radiative corrections. Most importantly, though, we are going to see that the
condition
(\ref{lprojbound}) introduces a more important problem: the scalar
mode becomes strongly coupled at unacceptably low energies.

 %%%%%%%%%%%%%%%%%%%%%%%%%%%%
\subsection{Strong coupling}
\label{sec:projstr}
 %%%%%%%%%%%%%%%%%%%%%%%%%%%%

In this section we extend the study of the projectable model beyond
the linear level and 
consider self-interaction of the scalar mode. As we will show, these
considerations rule out the model as a weakly coupled alternative to GR.
 
We will use the St\"uckelberg formalism described in
Sec.~\ref{sec:stuck}. To construct the appropriate khronon action one
notices \cite{bps1} that the projectable
model (\ref{ADMact1}) with the condition (\ref{pro}) is 
locally equivalent to the theory with the same action 
where the lapse $N$ is set to a constant; for example one can
choose\footnote{This choice breaks the time reparameterization
  invariance. As already mentioned, an alternative way to implement the
  projectability condition, which preserves the full FDiff symmetry,
  is to consider the $\alpha\to\infty$ limit of the non-projectable
  model. However, the approach adopted in this section allows to reach
our goal in a more direct manner.}
$N=1$. Indeed, the only difference between the two theories is in the
equation obtained from the variation of the action with respect to the
lapse. But this equation in the projectable model has the form of a
Hamiltonian constraint integrated over the whole space and does not
affect the local physics, cf. \cite{Mukohyama:2009mz}.   
In  
Ref.~\cite{bps1} 
the condition $N=1$
was
implemented by
introducing a Lagrange multiplier, and it was shown that the classical
low-energy dynamics of the khronon field are equivalent to that of a
pressureless fluid \cite{Mukohyama:2009mz}. 
In the present paper we adopt a different approach more
suitable for the analysis of the quantum properties of the system. 
Namely,
we implement the
constraint by adding to the action (\ref{ADMact1})
a
potential term 
\[
%\label{SM}
S_M=\int d^3x\,dt\sqrt{\gamma} N\,\frac{M^4}{4}\left(\frac{1}{N^2}-1\right)^2\;
\]
and considering the limit $M\to \infty$. In this limit the potential
becomes very steep and enforces $N=1$.
We will keep the
coefficient $M$ finite in the intermediate calculations 
and take the limit only at the last step.

Now we are ready to write down the covariant form of the action for
the projectable model. 
Using the prescription of Sec.~\ref{sec:stuck} we obtain,
\be
\label{Sphi}
S_{I}=\frac{1}{2}\int d^4x\sqrt{-g} \left[
-M_P^2\,{}^{(4)}R+\frac{M^4}{2}(X-1)^2
-\frac{M_\lambda^2}{X}
\left(\Box\varphi-\frac{\nabla^\mu\varphi\nabla^\nu\varphi}{X}
\nabla_\mu\nabla_\nu\varphi\right)^2\right]\;,
\ee
where $\varphi$ is the khronon field, $M_\lambda$ and $X$ are defined
respectively in (\ref{Mscales}) and (\ref{X}).
In this expression we have omitted the terms with higher
spatial derivatives from the potential (\ref{potential}). Note that
(\ref{Sphi}) coincides with the action of the ghost condensate model
\cite{ArkaniHamed:2003uy}. In particular, the last term in
(\ref{Sphi}) is a covariant realization of the term with four spatial
derivatives of the ghost condensate from
Ref.~\cite{ArkaniHamed:2003uy}. However, whereas in 
\cite{ArkaniHamed:2003uy} $M$ and $M_\lambda$ are assumed to be
of the same order, we are interested in the case $M_\lambda\ll M$.

To get a first feeling about the khronon dynamics it is instructive to
neglect the back-reaction of the khronon on the metric. Formally, this
is achieved by considering the  
limit
\be
\label{decoupling}
M_P\to\infty~,~~~~~M_\lambda,~M - \text{fixed}\;.
\ee
The
perturbations of the metric get frozen in this limit, so the khronon
dynamics effectively decouples from that of gravity. Following the
terminology adopted in massive gravity we 
refer to this regime as `decoupling
limit'. Note that the phenomenological constraint
(\ref{lprojbound}) implies $M_\lambda\ll M_P$ so that one of the
decoupling limit assumptions is automatically
satisfied. On the other hand the second assumption $M\ll M_P$ is in conflict
with 
the fact that we want to take $M$ eventually to infinity in order to
recover the projectable model. This means that the proper analysis of
the projectable model will require going beyond the previous decoupling
regime. 
However, working in the decoupling limit is still instructive. First,
it will allow to outline the steps of the analysis in a
simplified setting. 
Second, we will use the decoupling limit
results in the next section where we extend the model to a version of
the ghost condensation. 
With this in mind, let us proceed for the moment with the assumptions
(\ref{decoupling}).

Setting the metric perturbations to zero, $h_{\mu\nu}=0$, and writing
$\varphi=t+\chi$ we obtain:
\be
\label{Sphi1}
S_{I\chi}=\int d^4x\left[M^4\dot\chi^2-\frac{M_\lambda^2}{2}(\Delta\chi)^2-
M^4\,\dot\chi(\d_i\chi)^2
-M_\lambda^2\,\dot\chi
\big((\Delta\chi)^2+2\d_i\chi\d_i\Delta\chi\big)+\ldots\right]\;,
\ee
where we have written only the most relevant (cubic) interaction terms.
This  action describes a non-relativistic
scalar with derivative self-coupling.
The dispersion
relation for $\chi$ reads,
\be
\label{dispp}
\omega^2=\frac{M_\lambda^2}{2M^4}{\bf p}^4\;.
\ee
We can compare this dispersion relation with that obtained previously for the scalar mode of the
projectable model, Eq.~(\ref{disp}). 
In deriving (\ref{Sphi1}) we neglected the
higher-derivative terms. This explains the absence in (\ref{dispp}) of
terms with higher powers of momentum suppressed by $M_*$. Moreover, in
the projectable limit 
$M\to\infty$ 
(\ref{dispp}) reduces to 
$\omega^2=0$ which coincides with the $|\lambda-1|\to 0$ limit of
(\ref{disp}). The latter corresponds to the decoupling 
limit~(\ref{decoupling}). 

From the presence of derivative interactions in (\ref{Sphi1}) 
one expects  the field $\chi$ to become
strongly coupled above certain energy/momentum scale. The easiest
way to identify this scale is to perform power-counting. 
Due to the non-relativistic form of the dispersion
relation for the $\chi$-field the power counting rules in our case are 
different from the
standard ones. To identify the proper rules we follow the approach of
\cite{ArkaniHamed:2003uy,bps1} (see also
\cite{Polchinski:1992ed}). One notices that the quadratic part of
the action is invariant under the scaling transformations
\[
%\label{scalghost}
{\bf x}\mapsto b^{-1}{\bf x}~,~~~t\mapsto b^{-2}t~,~~~
\chi\mapsto b^{1/2}\chi\;.
\]
The interaction terms in (\ref{Sphi1}) scale as $b^{1/2}$ and
$b^{5/2}$ implying that the strength of the
interaction
grows at short distances. To simplify the further analysis let us choose
the 
units in such a way that
the 
quadratic part of the action takes canonical form, i.e. it contains
only order-one numerical coefficients. 
This is achieved by the rescaling  
\[
%\label{rescal1}
t=M^2M_\lambda^{-1}\hat t~,~~~~~~
\chi=M^{-1}M_\lambda^{-1/2}\hat\chi\;,
\]
which yields,
\[
%\label{Sphi2}
S_{I\chi}=\!\int\!\! d^3x d\hat t\left[\dot{\hat\chi}^2
-\frac{(\Delta\hat\chi)^2}{2}-
M M_\lambda^{-3/2}\dot{\hat\chi}(\d_i\hat\chi)^2
-M^{-3}M_\lambda^{1/2}\dot{\hat\chi}\big(
(\Delta\hat\chi)^2+2\d_i\hat\chi\d_i\Delta\hat\chi\big)+\ldots\right]\,.
\]
Now the momentum and frequency cutoffs (strong coupling scales) of the
theory are identified as the appropriate powers of the scale
suppressing the interaction terms,
\begin{align}
\label{cutp}
&\Lambda_{p,\;\mathrm{dec}}=\min\{M^{-2}M_\lambda^3,~M^{6/5}M_\lambda^{-1/5}\}\;,\\
%\label{cutomegahat}
&\Lambda_{\hat\omega,\;\mathrm{dec}}
=\min\{M^{-4}M_\lambda^6,~M^{12/5}M_\lambda^{-2/5}\}\;,\notag
\end{align}
where the subscript ``dec'' reminds that these expressions are
obtained in the decoupling limit.
Here $\hat\omega$ is the frequency corresponding to the rescaled time
$\hat t$; going back to the physical frequency one obtains
\be
\label{cutomega}
\Lambda_{\omega,\;\mathrm{dec}}=\min\{M^{-6}M_\lambda^7,~M^{2/5}M_\lambda^{3/5}\}\;.
\ee
Note that the cutoffs in momentum and energy are different reflecting
the non-relativistic nature of the theory; they are related by the
dispersion relation (\ref{dispp}).

In the projectable limit $M\to\infty$ both cutoffs
(\ref{cutp}), (\ref{cutomega}) vanish which would imply that the model
is strongly coupled at all scales. However, as already pointed above,
this reasoning has a caveat: the expressions (\ref{cutp}),
(\ref{cutomega}) are obtained under the assumptions $M_P\to\infty$ (which 
in particular implies $M\ll M_P$).
 We
are going to see that a \emph{finite} value of $M_P$ raises the cutoff from
zero; but it is still too low to be phenomenologically acceptable, 
cf.~\cite{Koyama:2009hc}. 

To obtain the correct value of the cutoff we have to take into account
the fluctuations of the metric. Let us first see how this affects the
analysis 
at the
quadratic level. Taking the metric perturbations in the form
(\ref{dec}) and expanding the action (\ref{Sphi}) to
quadratic order we obtain,
\be
\label{Sphi(2)}
\begin{split}
S_I^{(2)}=\int d^4x\bigg[\frac{M_P^2}{2}\big(-2\dot\psi^2
-2\psi\Delta\psi+4\phi\Delta\psi+4\psi\sqrt\Delta\dot B
+4\psi\ddot E\big)\\
+M^4(\dot\chi-\phi)^2
-\frac{M_\lambda^2}{2}\big(\sqrt\Delta B+\dot E+2\dot\psi
+\Delta\chi\big)^2\bigg]\;.
\end{split}
\ee  
Note that the first line here is nothing but the quadratic part
of the standard Einstein-Hilbert action. 
If one fixes in (\ref{Sphi(2)}) the gauge $\chi=0$ and takes the limit
$M\to\infty$ forcing $\phi$ to be constant one recovers the unitary
gauge Lagrangian\footnote{This procedure
does not reproduce the last two terms in (\ref{L2proj}). These
higher-derivative terms have been
omitted in the low-energy action (\ref{Sphi(2)}).} (\ref{L2proj}).
For our present purposes it is more convenient  
to choose instead the gauge
\[
B=0~,~~~2\psi+E=0\;,
\] 
where khronon perturbations do not vanish.
To integrate out the non-propagating degrees of freedom recall that we
are working in the regime 
$M_\lambda\ll M_P$. From the dispersion relation (\ref{disp}) we know
that in this case the frequency of scalar perturbations is much
smaller than the spatial momentum, 
\be
\label{frmom}
\omega\sim\frac{M_\lambda}{M_P}\,|\bf p| \ll |\bf p|\;.
\ee 
Thus we can neglect the term $\dot\psi^2$ compared to $\psi\Delta\psi$
in (\ref{Sphi(2)}). Then the
equation of motion of $\psi$ 
implies $\psi=\phi$; substituting this into the action we obtain, 
\[
%\label{Sphi(2)1}
S_I^{(2)}=\int d^4x\bigg[M_P^2\;\phi\Delta\phi+M^4(\dot\chi-\phi)^2-
\frac{M_\lambda^2}{2}(\Delta\chi)^2\bigg]\;.
\]
Varying with respect to $\phi$ one finds
\be
\label{phichi}
\phi=\frac{M^4\dot\chi}{M_P^2\Delta+M^4}\;,
\ee
and hence
\be
\label{Sphi4}
S_{I}^{(2)}=\int
d^4x\bigg[\frac{M_P^2M^4}{M_P^2\Delta+M^4}\dot\chi\Delta\dot\chi
-\frac{M_\lambda^2}{2}(\Delta\chi)^2\bigg]\;.
\ee
This gives the dispersion relation
\be
\label{dispp1}
\omega^2=-\frac{M_\lambda^2}{2M_P^2}{\bf
  p}^2+\frac{M_\lambda^2}{2M^4}{\bf p}^4\;,
\ee
which reduces to (\ref{dispp}) in the decoupling limit
$M_P\to\infty$. On the other hand, in the projectable limit
$M\to\infty$ it correctly reproduces the first term in the exact
dispersion relation (\ref{disp}).

Let us turn to the interactions. One has to compare contributions
coming from various terms in the action (\ref{Sphi}). We start by
considering the second term in (\ref{Sphi}). Due to the inequality
(\ref{frmom}) the leading interaction is given by the cubic term with
the smallest number of time derivatives. A simple calculation yields
\begin{align}
S_M^{(3)}
&=\int d^4x\big[-M^4(\dot\chi-\phi)(\d_i\chi)^2+\ldots\big]\notag\\
&=\int
d^4x\bigg[
-\frac{M_P^2M^4}{M_P^2\Delta+M^4}\Delta\dot\chi(\d_i\chi)^2
+\ldots\bigg]\;,
\label{Sphi3}
\end{align}
where in the second line we substituted $\phi$ from (\ref{phichi}).
Note that in the decoupling limit this interaction term reproduces the
third term in the action
(\ref{Sphi1}). Next, let us estimate the contributions from the first
and third terms in (\ref{Sphi}) and show that they are suppressed
compared (\ref{Sphi3}). This can be done without explicit calculation
on purely dimensional grounds. Indeed, $^{(4)}R$ does not explicitly
depend on the khronon and has mass dimension two. Thus the possible
leading contribution (the one without time derivatives when written in
terms of the metric) 
has the schematic form
\[
M_P^2\sqrt{-g} ~^{(4)}R\sim M_P^2\phi^2\Delta\phi
\sim M_P^2\dot\chi^2\Delta\dot\chi\;,
\]
where in the last relation we used Eq.~(\ref{phichi}) and took the
limit $M\to\infty$. This contains more time derivatives than
(\ref{Sphi3}) and hence is suppressed. Finally, the schematic form of
the leading contribution from the third term in (\ref{Sphi}) is 
$M_\lambda^2\dot\chi(\Delta\chi)^2$. This is clearly suppressed
compared to (\ref{Sphi3}) by the ratio $(M_\lambda/M_P)^2$.

In the projectable limit
$M\to\infty$ both the quadratic action 
(\ref{Sphi4}) and the interaction term (\ref{Sphi3})
simplify. Combining them together we obtain
\[
%\label{Sphi5}
S_{I}=\int
d^4x\bigg[M_P^2\dot\chi\Delta\dot\chi
-\frac{M_\lambda^2}{2}(\Delta\chi)^2-
M_P^2\Delta\dot\chi(\d_i\chi)^2
+\ldots\bigg]\;.
\]
It is now straightforward to find the strong coupling scales of the
model. Upon bringing the quadratic part of the action into canonical
form by the rescaling 
\[
%\label{rescal2}
t=M_PM_\lambda^{-1}\hat t~,~~~~~
\chi=M_P^{-1/2}M_\lambda^{-1/2}\hat\chi\;,
\]
one finds that the interaction term is suppressed by the parameter
\[
%\label{cutptrue}
\Lambda_p=M_\lambda^{3/2}M_P^{-1/2}\;,
\]
which thus sets the momentum cutoff of the theory. The energy cutoff
contains additional factor $M_\lambda M_P^{-1}$ and reads
\[
%\label{cutomegatrue}
\Lambda_\omega=M_\lambda^{5/2}M_P^{-3/2}\;.
\]
Note that these cutoffs go down to zero in the naive GR limit
$\lambda\to 1$ ($M_\lambda\to 0$). This agrees with the results of
\cite{Koyama:2009hc}.

Using the definition of $M_\lambda$, Eq.~(\ref{Mscales}),
we find that the stability bound
(\ref{lprojbound}) implies
\be
\label{Lpp}
\Lambda_p\lesssim 10^{-17}\,\mathrm{eV}\;.
\ee
This corresponds to the distance of order $10^{13}\,$cm, within which 
the theory is strongly coupled and the perturbative analysis breaks
down. Note that the strong coupling cannot be resolved by the
higher-derivative terms: such terms would lead to sizable
modifications of the Newton's law that are forbidden at these
distances. We conclude that the projectable model fails to provide
adequate description of gravity within distances 
$\sim 10^{13}\,$cm. 

%\newpage

%%%%%%%%%%%%%%%%%%%%%%%%%%%%%%%%%%%%%%%
\section{Extension \`a la ghost condensation}
\label{sec:ghc}
%%%%%%%%%%%%%%%%%%%%%%%%%%%%%%%%%%%%%%%

The analysis of the previous section suggests a possible way to
address the problems of the projectable model. The idea is to consider
a modification of the theory with finite value of the parameter $M$ in
the khronon action (\ref{Sphi}) \cite{bps1}. 
This approach implies reducing the symmetry of the theory
from FDiffs down to RFDiffs as the khronon action is no longer
invariant under the general reparameterizations (\ref{phirepar}) but
only under the shifts (\ref{shift}). We presently explore this option
and show that in spite of the possibility to improve the behavior of
the khronon at low energies  
the theory still encounters serious problems in the UV
completion.

As already pointed out, at finite values of $M$ the low-energy 
khronon action
(\ref{Sphi})
coincides with the action of the ghost condensate model
\cite{ArkaniHamed:2003uy}. The latter is a consistent effective theory
describing low-energy modification of GR. 
From the EFT point of view it is natural to
assume that 
the scales $M$ and $M_\lambda$ appearing in (\ref{Sphi}) are of
the same order. Then
both the momentum and frequency cutoffs obtained in the decoupling limit
(\ref{cutp}), (\ref{cutomega}) are finite and of order $M$. 
From the dispersion
relation (\ref{dispp1}) one observes that the instability of the
scalar graviton at low values of momenta is still present in this
model but is truncated by the quartic term  
at 
\[
%\label{pcr}
p\sim M^2/M_P\;.
\]
This corresponds to the maximal instability rate
\[
%\label{instghost}
\Im \omega\sim M^3/M_P^2\;.
\]
This rate is 
slower than the present rate of Hubble expansion for 
\be
\label{Mghc}
M\lesssim 10\;\mathrm{MeV},
\ee 
which would make the
instability harmless.\footnote{It is argued in \cite{ArkaniHamed:2005gu} that
  nonlinear dynamics of the ghost condensate suppresses the
  exponential growth of perturbations predicted by the linear theory,
  making the model phenomenologically viable for $M$
  up to $100\;\mathrm{GeV}$.}

The problems appear, however, when one tries to UV complete the model to
a renormalizable theory of gravity. 
First, a necessary requirement to obtain a weakly coupled UV
completion is that the strong coupling observed within low-energy EFT
must be resolved by the higher-derivative operators,
cf.~\cite{bps3}. This implies that the scale $M_*$ suppressing higher
derivatives should be lower than $M$, $M_*<M$. Thus we obtain
\be
\label{M*ghc}
M_*\lesssim 10\;\mathrm{MeV}\;.
\ee
Note that this
is many orders of magnitude smaller than $M_P$ so that
the model contains two very different scales. For the gravitational
sector alone, this seems to be only an esthetic problem: 
the values (\ref{M*ghc}) are compatible with the direct
gravitational bound (\ref{M*bound}).
Besides, 
we expect this hierarchy to
be stable under radiative corrections.
The
scale which effectively truncates the power-law divergencies in the
model is the lower scale $M_*$; therefore, the corrections to
both $M_P$ and $M_*$ are small\footnote{See also the discussion of
  a similar issue in the next section.}.
   However, the real tension arises when
we take into account coupling of the theory to matter. The size of the
higher-derivative Lorentz violating terms in the action of the matter
sector is
experimentally constrained from below. 
The leading effect of these terms is the
modification of dispersion relations of the matter fields at high
energies due to contributions with 
higher powers of spatial momentum, see Eq.~(\ref{dispmat1}) below. 
The observational lower bound on the scale
$M_*^{(mat)}$ suppressing these contributions, Eq.~(\ref{M*mat1}), is
much higher than 10 MeV.
On general grounds one expects the scale $M_*^{(mat)}$ to enter into
radiative corrections for $M_*$ thus giving rise to a fine-tuning
problem\footnote{The fine tuning may be absent if there is a symmetry
protecting $M_*$ from this kind of corrections.}. 
A more detailed study of this issue is beyond the scope of
the present article.

The second problem stems from the reduction of the symmetry from
FDiffs down to RFDiffs. To get a renormalizable
theory one must consider the most general RFDiff-invariant 
action (\ref{ADMact3}) which contains,
among
other terms, the term with time derivatives of the lapse,\footnote{Even if
  this term is absent in the bare action, one expects it to be
  generated by quantum corrections.}  
\be
\label{SdotN}
S_{\lambda_1}= \frac{ M_{\lambda_1}^2}{2}\int \di^3 x\, \di t\,
\sqrt{\gamma}N~(\dot N-N^i\pd_i N)^2\;,
\ee 
where $M_{\lambda_1}$ is defined in (\ref{Mscales}).
The easiest way to see that this term leads to pathologies
is to adopt
the St\"uckelberg picture. The covariant form of  
(\ref{SdotN}) reads
\be
\label{SdotN1}
S_{\lambda_1}=\frac{M_{\lambda_1}^2}{2}\int d^4 x\sqrt{-g}~
\frac{\left({\nabla^\mu\varphi\,\nabla^\nu\varphi\,
\nabla_\mu\nabla_\nu\varphi}\right)^2}{X^{2}}\;.
\ee
Let us work in the decoupling limit; the conditions for the validity
of this approximation will be specified below.
One writes $\varphi=t+\chi$ and expands
(\ref{SdotN1}) up to quadratic order in~$\chi$:
\[
%\label{Sdot2}
S_{\lambda_1}=\frac{M^2_{\lambda_1}}{2}\int d^4x\; (\ddot\chi)^2\;.
\] 
This contains four time derivatives of the khronon perturbations
which implies the presence of a second helicity-0 mode besides the
excitation 
studied so far. As the new mode appears due to
higher time derivatives one is tempted to qualify it as a ghost. In
fact, we are going to show that this mode also exhibits a gradient
instability, so it is more appropriate to call it {\em tachyonic
  ghost}. In Appendix~\ref{app:B} we discuss 
the difference between
the tachyonic ghost and  an ordinary
 ghost with stable dispersion relation\footnote{Of course, both type of ghosts 
signal pathologies of the
  theory.}.   
In particular, we demonstrate
there that the presence of the
tachyonic ghost is compatible with the fact that in
the unitary gauge the action for the model,
Eq.~(\ref{ADMact3}), is only second order in time derivatives.

To proceed we need the quadratic action for the khronon
perturbations. This is obtained in the usual way by first
covariantizing the terms entering (\ref{ADMact3}), (\ref{potIII}) and
then taking the decoupling limit. This yields,
\be
\label{SchiIII}
S_{III\chi}=\int d^4x \left[\frac{M^2_{\lambda_1}}{2}(\ddot\chi)^2
+M^4\dot\chi^2-\frac{M_\lambda^2}{2}(\Delta\chi)^2
+\frac{M_\alpha^2}{2}(\d_i\dot\chi)^2\right]\;.
\ee  
In deriving this expression we have omitted the contributions of terms
with higher spatial derivatives; thus the domain of validity of
(\ref{SchiIII}) is restricted to spatial momenta smaller than $M_*$. 
The last term in (\ref{SchiIII}) represents the contribution of the 
term $\alpha a_ia^i$ in the
potential (\ref{potIII}) (recall the definition of
$M_\alpha$ in (\ref{Mscales})). 
The action (\ref{SchiIII}) describes two modes with 
dispersion relations
\be
\label{dispIII}
\omega^2=-\left(\frac{M^4}{M_{\lambda_1}^2}
+\frac{M^2_\alpha}{2M_{\lambda_1}^2}{\bf p}^2\right)
\pm\sqrt{\left(\frac{M^4}{M_{\lambda_1}^2}
+\frac{M^2_\alpha}{2M_{\lambda_1}^2}{\bf p}^2\right)^2
+\frac{M^2_\lambda}{M_{\lambda_1}^2}{\bf p}^4}\;.
\ee
At small values of momenta we find,
\begin{align}
\omega^2_{old}&= 
\frac{M_\lambda^2}{2M^4} {\bf p}^4 + \dots 
%\label{lambda1a}
\notag\\  
\omega^2_{new}&= -  \frac{2 M^4}{M_{\lambda_1}^2} -
\frac{M_\alpha^2}{M_{\lambda_1}^2} {\bf p}^2   
-\frac{M_\lambda^2}{2M^4} {\bf p}^4 +  \dots ~.
\label{lambda1}
\end{align}
In the first expression
one recognizes the dispersion relation (\ref{dispp}) for the 
previously encountered gapless mode. The second mode (\ref{lambda1})
is unstable.
It 
possesses a frequency gap of order $M^2/M_{\lambda_1}$.
For the choice
$M_{\lambda_1}\sim M$, natural from the low-energy effective theory
point of view, the frequency of this mode lies above the scale $M$.
Thus this mode would be simply discarded as unphysical in the EFT
considerations with cutoff $M$, such as in
\cite{ArkaniHamed:2003uy}. However, our case is different: we are
looking for a UV-complete model and thus have to take into account
all possible excitations of the system.

According to the above expressions the instability rate of the new mode
grows with momentum, the fastest instability occurring at 
$|{\bf p}|\sim M_*$. [At larger momenta terms with higher spatial
derivatives in the full action (\ref{ADMact3}) can, 
in principle, stabilize the mode.] A lower estimate for the
instability rate is obtained by keeping only 
the last term under the square root
in Eq.~(\ref{dispIII}). This yields,
\be
\label{instrate}
\Im\, \omega_{new} \gtrsim M_*^{3/2}M_{\lambda_1}^{-1/2}\;,
\ee
where we have assumed
\be
\label{4Ms}
M\sim M_\alpha\sim M_\lambda\gtrsim M_*\;.
\ee
Requiring the instability 
to be slower than the Hubble rate $H_0$ and taking into
account the experimental bound (\ref{M*bound}) on $M_*$ one
obtains,
\be
\label{Mhuge}
M_{\lambda_1}\gtrsim 10^{60}\,\mathrm{eV}\approx 10^{32}M_P\;.
\ee
Having such a huge value for $M_{\lambda_1}$
is rather
unsatisfactory. First, it means that one introduces into the theory, 
besides $M_*$ and
$M_P$,
one more
hierarchical scale
$M_{\lambda_1}\gg M_P$. We do not know whether this
hierarchy is technically natural or not. 
More importantly, this term also gives rise to interactions
which are enhanced by the large value of $M_{\lambda_1}$. 
We will demonstrate shortly that this
reintroduces the strong coupling problem.

Before addressing the interactions let us make a
step back and discuss
the conditions for the validity of the decoupling limit. The simplest
requirement would be that all the parameters in the khronon action
(\ref{SchiIII}) are smaller than the Planck mass. This is
automatically satisfied by the parameters $M_\lambda$, $M$, $M_\alpha$
that must be small for the stability of the gapless (ghost-condensate)
mode. However, the value (\ref{Mhuge}) of $M_{\lambda_1}$ clearly
violates this condition. Thus additional considerations are
needed to establish whether the decoupling limit holds or not for
$M_{\lambda_1}$ as large as (\ref{Mhuge}). 
A rigorous 
  method to do this is to verify 
that the terms in the Lagrangian describing mixing between
  the khronon field and the metric components are small compared to
  the other terms. Instead of following this route we will take a
  shortcut and
compare the decoupling limit dispersion relations (\ref{dispIII}) for
the scalar modes with the exact expressions; their coincidence will serve
as a criterion for decoupling. The exact dispersion relations 
are obtained in Appendix~\ref{app:B} within the unitary
gauge, see Eq.~(\ref{QFdisp}). In the case $M_{\lambda_1}\gg M_P$ they
simplify to\footnote{\label{foot}Note that these expressions immediately imply a
  lower estimate for the
instability rate of the new mode which is complementary to
(\ref{instrate}), 
\[
%\label{disppIII}
\Im \omega_{new}>\frac{M_\lambda}{\sqrt{2} M_P}|{\bf p}|\;.
\]
This estimate does not depend on
$M_{\lambda_1}$.
Evaluating it at $|{\bf p}|\sim
M_*$ and requiring the instability to be
slower than the Universe expansion rate we obtain the upper bound
\be
\label{Mlamup}
M_\lambda\lesssim 0.1\,\mathrm{eV}\;.
\ee
This is marginally compatible with the requirement (\ref{4Ms}) and the
experimental bound (\ref{M*bound}).},
\[
%\label{dispexIII}
\omega^2=\bigg(-\frac{M_{\lambda}^2}{4M_P^2}\pm
\sqrt{\frac{M_{\lambda}^4}{16M_P^4}
+\frac{M_{\lambda}^2}{M_{\lambda_1}^2}}\bigg)\,{\bf p}^2\;.
\]
On the other hand, the decoupling limit result (\ref{dispIII}) at
large $M_{\lambda_1}$ becomes
\be
\label{dispdecIII}
\omega^2=\pm\frac{M_\lambda}{M_{\lambda_1}}{\bf p}^2\;.
\ee
The two expressions coincide and thus the decoupling holds, provided
\be
\label{M1Dec}
M_{\lambda_1}\ll\frac{M_P^2}{M_\lambda}\;.
\ee
Note that this allows for $M_{\lambda_1}$ to be much
larger than $M_P$.  Taking $M_\lambda\sim 0.1\,\mathrm{eV}$ as in the
previous estimates
(see also the footnote \ref{foot}) one finds that the decoupling limit
is applicable up to $M_{\lambda_1}$ of order~(\ref{Mhuge}).

The previous considerations show that to estimate the scale of strong
coupling we can concentrate on the self-interaction of the khronon
neglecting its mixing with the metric. Clearly, for large
$M_{\lambda_1}$ the leading interactions come from the term
(\ref{SdotN1}). At the cubic level we obtain the following
contribution,
\[
%\label{SdotN2}
S_{\lambda_1}^{(3)}=-2M^2_{\lambda_1}\int d^4x\;
\ddot\chi\d_i\dot\chi\d_i\chi\;.
\]
This must be combined with the quadratic action (\ref{SchiIII}). 
The form of the dispersion relation
(\ref{dispdecIII}) shows that at large $M_{\lambda_1}$ 
the latter action 
is dominated  by the first and the third
terms. Retaining only these two terms we obtain the khronon action in the
limit of interest
\[
S_{III\chi}=\int d^4x\bigg[\frac{M^2_{\lambda_1}}{2}(\ddot\chi)^2-
\frac{M^2_\lambda}{2}(\Delta\chi)^2
-2M^2_{\lambda_1}\ddot\chi\d_i\dot\chi\d_i\chi+\ldots\bigg]\;.
\]
The quadratic part is
brought to canonical form by the rescaling 
\[
t=M_\lambda^{-1/2}M_{\lambda_1}^{1/2}\,\hat t~,~~~~
\chi=M_\lambda^{-3/4}M_{\lambda_1}^{-1/4}\,\hat \chi\;.
\]
From the scale suppressing the interaction term in the resulting
action one reads off the momentum and frequency cutoffs,
\[
\Lambda_p=M_\lambda^{5/4}M^{-1/4}_{\lambda_1}~,~~~~
\Lambda_\omega=M_\lambda^{7/4}M^{-3/4}_{\lambda_1}\;.
\]
Substituting here $M_\lambda$ from (\ref{Mlamup}) and $M_{\lambda_1}$
from (\ref{Mhuge}) one obtains\footnote{
Note that this is close to
  the result in the projectable case, Eq.~(\ref{Lpp}).
} 
$\Lambda_p\lesssim
10^{-16}\,\mathrm{eV}$. Such a low cutoff is phenomenologically unacceptable.
This shows that the initial hope, 
namely that the model could be UV complete and weakly coupled, is not met in 
reality\footnote{One may wonder if it is possible to cure the model by
  choosing some of the parameters in the action (\ref{SchiIII})
  negative. A simple reasoning demonstrates that this is not the
  case. The analysis in the unitary gauge (Appendix~\ref{app:B}) shows
  that negative values of $M_{\lambda_1}^2$, $M_\lambda^2$ imply
  negative energies in the UV. Next, making $M_\alpha^2$ negative does
not essentially improve the behavior of the tachyonic ghost. Finally,
the choice $M^4<0$, while removing the instability from the mode
(\ref{lambda1}), destabilizes the old gapless mode. This brings us
back to the situation studied in the previous section.}.

The general conclusion of this section is that 
reducing the symmetry of the theory from FDiff to RFDiffs cannot 
possibly improve the scalar sector. 
The reason is that the smaller symmetry allows for the new operators
such as \eqref{SdotN}  
which bring in additional difficulties in the form of tachyonic
ghosts.
%In passing, note that 
%the same problem appears to be inevitable also 
%in other khrono-metric models with even less symmetry, such
%as the case (iii) listed in the introduction  
%(although the same conclusion does not necessarily apply to 
%the cases (iv) and (v)), at least without additional ingredients. 
In retrospect, we learn that the symmetry under the FDiffs 
(or the khronon reparameterizations \eqref{phirepar} in covariant language) 
plays the quite important role of preventing this kind of pathologies.
In the next section, we return to the other model compatible with the
FDiffs ---
the general non-projectable case --- where 
this problem is automatically turned away.\footnote{
Clearly, the symmetry \eqref{phirepar} must be non-anomalous 
in order that this statement remains valid in the quantum theory. 
This question will not be addressed here.}

%%%%%%%%%%%%%%%%%%%%%%%%%%%%
\section{The healthy extension}
\label{sec:healthy}
%%%%%%%%%%%%%%%%%%%%%%%%%%%%

%%%%%%%%%%%%%%%%%%%%%%%%%%%%
\subsection{Stability and absence of strong coupling}
\label{sec:healthy1}
%%%%%%%%%%%%%%%%%%%%%%%%%%%%

We presently consider the non-projectable FDiff-invariant model with
the action (\ref{ADMact2}), (\ref{potadd}). We are going to show that,
remarkably,  
the problems
associated to the additional modes which plague the two models
considered previously are absent in this case
\cite{bps2,bps3}.  

Let us start by studying the spectrum of linear perturbations around flat
background. As usual, we concentrate on the scalar sector and use the
decomposition (\ref{dec}). In the unitary gauge 
the quadratic Lagrangian reads
\be
\label{L2imp}
\begin{split}
&{\cal L}^{(2)}=\frac{M_P^2}{2}\bigg[-2\dot\psi^2
-2\psi\Delta\psi+4\phi\Delta\psi
+4\psi\sqrt\Delta\dot B +4\psi\ddot E-(\lambda-1)\left(\sqrt\Delta B+
\dot E+2\dot\psi\right)^2\\
&+\alpha(\d_i\phi)^2
-\frac{f_1}{M_*^2}(\Delta\psi)^2-\frac{2f_2}{M_*^2}\Delta\phi\Delta\psi
-\frac{f_3}{M_*^2}(\Delta\phi)^2-\frac{g_1}{M_*^4}\psi\Delta^3\psi
-\frac{2g_2}{M_*^4}\phi\Delta^3\psi
-\frac{g_3}{M_*^4}\phi\Delta^3\phi\bigg]\;,
\end{split}
\ee
where $f_n$, $g_n$ are related to the coefficients in front of the
higher derivative operators in the potential (\ref{potadd}); the
precise form of this relation is not important for our
purposes. Integrating out the non-dynamical fields $B$, $E$ and $\phi$
we obtain,
\be
\label{Lscalar}
\begin{split}
{\cal L}^{(2)}_{II}=\frac{M_P^2}{2}\bigg\{
\frac{2(3\lambda-1)}{\lambda-1}\dot\psi^2
+\psi\,\frac{P[M_*^{-2}\Delta]}{Q[M_*^{-2}\Delta]}\Delta \psi\bigg\}\;,
\end{split}
\ee
where the polynomials $P$, $Q$ have the form,
\begin{align}
P[x]=&(g^2_2-g_1g_3)x^4-(g_1f_3+g_3f_1-2g_2f_2)x^3
+(f_2^2-4g_2-f_1f_3-2g_3-g_1\alpha)x^2\notag\\
&-(2f_3+f_1\alpha+4f_2)x+(4-2\alpha)\;,
\label{poly1}\\
\label{poly2}
Q[x]=&g_3x^2+f_3x+\alpha\;.
\end{align}
The Lagrangian (\ref{Lscalar}) describes a single mode which is free
of pathologies provided that 
two conditions
are satisfied. First, the positivity of the 
kinetic term can be achieved by choosing\footnote{We do not consider
  the other option $\lambda<1/3$ because it corresponds to a strong
  deviation from GR, unacceptable from the phenomenological
  viewpoint.} 
$\lambda>1$. 
Second, from  the dispersion
relation of the propagating mode $\psi$,
\be
\label{dispgood}
\omega^2=\frac{\lambda-1}{2(3\lambda-1)}\;
\frac{P[-{\bf p}^2/M_*^2]}{Q[-{\bf p}^2/M_*^2]}\;{\bf p}^2\;,
\ee
one reads off the condition to avoid exponential 
instabilities,
\be
\label{condPQ}
P[x]/Q[x]>0~~~~~~ \mathrm{ at}~~ x<0\;.
\ee
This puts certain restrictions on the coefficients $\alpha$,
$f_n$, $g_n$, that are presented in Appendix~\ref{app:C}. 
In particular, at low energies the dispersion relation
takes the form,
\be
\label{dispimp}
\omega^2=\frac{\lambda-1}{3\lambda-1}\left(\frac{2}{\alpha}-1\right)
{\bf p}^2
\;.
\ee
Thus stability requires\footnote{As discussed in
  Sec.~\ref{sec:theories}, by taking the limit $\alpha\to\infty$ in
  the model considered here one obtains the
  projectable Ho\v rava's model. Clearly, the corresponding 
values of $\alpha$ are
outside the stability range (\ref{alphagood}), so we again find that
the projectable model is unstable.}
\be
\label{alphagood}
0<\alpha<2\;.
\ee
Note that the dispersion relation (\ref{dispimp}) describes a gapless
mode propagating with constant velocity which is generically different
from one (the velocity of
the helicity-2 modes, i.e. gravitons). This signals that in the model
at hand Lorentz symmetry is broken down to arbitrary low
energies. In the UV the dispersion relation (\ref{dispgood}) takes the
form $\omega^2\propto {\bf p}^6$ which obeys the anisotropic scaling
with $z=3$. This is compatible with the power-counting arguments in
favor of renormalizability. 

It is important to realize that the healthy behavior of the scalar mode
can be achieved simultaneously with the stability in the sector of the
helicity-2 perturbations. 
Consider operators in the action (\ref{ADMact2}),
(\ref{potadd}) which contribute at the quadratic level. Upon integrating 
by parts and using Bianchi identities 
one obtains a list of $10$ inequivalent
combinations, 
%\bseq
%\label{oper}
\begin{align}
%\label{oper1}
&(\mathrm{dim}~2)~~~~ R,~a_ia^i\;,\notag\\
%\label{oper2}
&(\mathrm{dim}~4)~~~~ R_{ij}R^{ij},~R^2,~
R\nabla_i a^i,~a_i\Delta a^i\;,\notag\\
%\label{oper3}
&(\mathrm{dim}~6)~~~~ 
(\nabla_iR_{jk})^2,~(\nabla_i R)^2,~\Delta R\, \nabla_i a^i,~
a_i \Delta^2 a^i\;.\notag
\end{align}
%\eseq  
The dispersion relation of the
helicity-2 modes depends only on the coefficients in front of the 
operators in the first
column of this list. After fixing these to
ensure stability of the helicity-2 modes, we
still have the freedom to choose the coefficients of the remaining 
operators in the list. The number of free parameters matches with the
number of coefficients 
$\alpha$, $f_n$, $g_n$ in the scalar Lagrangian implying that we have 
freedom to adjust the latter coefficients
to satisfy~(\ref{condPQ}). 

To get more insight into 
the dynamics of the model (in particular, at the nonlinear level), we
make use of the St\"uckelberg formalism. 
The covariant form of the model action was given
before, see (\ref{covar2}). It is convenient to rewrite it as
\be
\label{covar21}
S_{II}=-\frac{M_P^2}{2}\int \di^4x\sqrt{-g}\Big\{{}^{(4)}R
+(\lambda-1)(\nabla_\mu u^\mu)^2
+\alpha\, u^\mu u^\nu\nabla_\mu u^\rho\nabla_\nu u_\rho
+\ldots\Big\}\;, 
\ee 
where $u_\mu$ is related to the khronon field $\varphi$ by
(\ref{umu}) and dots stand for the
terms with higher derivatives which are not important at low energies.
The action (\ref{covar21}) is closely related
\cite{bps3,Jacobson:2010mx} to the action of the Einstein-aether theory
\cite{Jacobson:2000xp} (see 
\cite{Jacobson:2008aj} for a review). The latter is an effective
theory describing breaking of Lorentz-invariance by a time-like vector
field with unit norm. The difference between our case and the 
Einstein-aether theory is that in our model the unit vector $u_\mu$ is
by construction hypersurface orthogonal, i.e. it is completely
characterized by the scalar khronon field $\varphi$. The similarity
between (\ref{covar21}) and the Einstein-aether theory will be exploited
in the next subsections where we study phenomenological consequences of
the model. 

Let us use
(\ref{covar21}) 
to study interactions of the khronon perturbations
at low energies. It is convenient to introduce the scales
$M_\lambda$ and $M_\alpha$ as defined in (\ref{Mscales}). These scales 
characterize the khronon action. We will assume them to be 
much smaller than $M_P$, so
that the metric perturbations are frozen out (in other words, we will be
working in the decoupling limit). This assumption is justified by the
phenomenological bounds (\ref{PPNbound}) that will be obtained in 
Sec.~\ref{sec:healthy3}, and which 
constrain the dimensionless parameters $|\lambda-1|$, $\alpha$ to be much
smaller than one\footnote{Note
 though
 that the bounds (\ref{PPNbound}) are much
 weaker than the analogous bounds for the models of 
the previous sections, cf. (\ref{lprojbound}),
(\ref{Mghc}).}.
Writing down $\varphi=t+\chi$ and expanding (\ref{covar21}) up to
cubic order in $\chi$ we obtain,
\begin{align}
\label{SIIchi}
S_{II\chi}=\int d^4 x
\bigg[
\frac{M_\alpha^2}{2}(\d_i\dot\chi)^2
-\frac{M_\lambda^2}{2}(\Delta\chi)^2 &-M_\lambda^2 \;\dot\chi\,
\Big((\Delta\chi)^2+2\d_i\chi\d_i\Delta\chi\Big)\notag\\
&+ M_\alpha^2 \; \Big( \dot \chi\d_i\ddot\chi\d_i\chi
-\d_i\dot\chi  \d_j\chi \d_i\d_j\chi
\Big)+\ldots
\bigg]\;.
\end{align}
Let us analyze this expression. The action
(\ref{SIIchi}) describes a propagating mode with dispersion relation 
\be
\label{dispchi}
\omega^2=\frac{M_\lambda^2}{M_\alpha^2}{\bf p}^2\;.
\ee 
This coincides with the exact dispersion relation (\ref{dispimp}) in
the limit $|\lambda-1|,~\alpha\ll 1$.
The form of the action (\ref{SIIchi}) is uniquely fixed by the 
reparameterization
symmetry (\ref{phirepar}) of the khronon and by the Lorentz symmetry,
both non-linearly realized on the khronon perturbation $\chi$. 
Up to quadratic order
the
reparameterization transformations read
\[
%\label{phireparpert}
\chi\mapsto  \chi+\epsilon(t)+\dot\epsilon(t)\chi+\ldots\;.
\]
where $\epsilon(t)$ is an arbitrary function of time. Under boosts the
field $\chi$ transforms as 
\[
%\label{Lorchi}
\chi\mapsto\chi+\varepsilon_i x_i+\varepsilon_i x_i\dot\chi
+\varepsilon_i t \d_i\chi+\ldots\;,
\] 
where $\varepsilon_i$ is a 3-dimensional vector characterizing the
boost. It is straightforward 
to check that (\ref{SIIchi}) is the only action invariant under these
symmetries up to cubic order. 
Finally, one observes that (\ref{SIIchi})
contains an interaction term with three time derivatives; it
produces a contribution with third
time derivative in the equation of 
motion for $\chi$. One may be worried that this leads to appearance of
a new unwanted
degree of freedom. However, from the analysis in the unitary gauge, we know
that this degree of freedom is spurious and it should be possible to eliminate
it with an appropriate choice of variables (which corresponds to
fixing the foliation consistently). To see how this is done
explicitly,  
one considers the change of variable
\[
%\label{chichange}
\chi=\tilde\chi+\tilde\chi\dot{\tilde\chi}\;.
\]   
This substitution
  can be interpreted as the change of the time foliation, $t\mapsto
  t-\chi$. Indeed, up to cubic order
$\tilde\chi(t,{\bf x})=\chi(t-\chi,{\bf x})$.
In terms of the new variable
the action takes the form
\begin{align}
\label{SIIchitilde}
S_{II\chi}=\int d^4 x
\bigg[
\frac{M_\alpha^2}{2}(\d_i\dot{\tilde\chi})^2
&-\frac{M_\lambda^2}{2}(\Delta\tilde\chi)^2 -M_\lambda^2 \;\tilde\chi\,
\Delta\tilde\chi\,\Delta\dot{\tilde\chi}\notag\\
&+ M_\alpha^2 \; \bigg(\frac{1}{2} \dot{\tilde\chi}(\d_i\dot{\tilde\chi})^2
-\d_i\dot{\tilde\chi}  \d_j\tilde\chi \d_i\d_j\tilde\chi
\bigg)+\ldots
\bigg]\;.
\end{align}
We observe that a term with three time derivatives is still present but its
structure has changed: its contribution to the equation of motion
contains now only second time derivative. One concludes that no new
degrees of freedom appear. It is worth emphasizing that 
the existence of the change of
variables with the above properties follows from the general statement 
proved in Appendix~\ref{app:A} that the equations of
motion for the khronon field are second order in time for an
appropriate choice of the coordinate system. 

The low-energy action (\ref{SIIchitilde}) contains derivative couplings
which become stronger as the energy / momentum grows. Let us estimate
the scale where the strength of the interactions would become of order
one. Performing the rescaling
\[
%\label{thealthy}
t=M_\alpha  M_\lambda^{-1}\;\hat t~,~~~~~~
\tilde\chi=M_\alpha^{-1/2}M_\lambda^{-1/2}\;\hat\chi\;,
\]
which casts the quadratic part of the action into canonical form we obtain,
\begin{align}
\label{Shealthy2}
S_{II\chi}=\int \di^3 x\, \di\hat t\;
\bigg[
\frac{(\d_i\dot{\hat\chi})^2}{2}
&-\frac{(\Delta\hat\chi)^2}{2}
-\frac{M_\lambda^{1/2}}{M_\alpha^{3/2}} \;  
\hat\chi \, \Delta\hat\chi \Delta\dot{\hat\chi}
\notag\\
&+\frac{M_\lambda^{1/2}}{2 M_\alpha^{3/2}} 
\dot{\hat\chi}(\d_i\dot{\hat\chi})^2
-\frac{M_\alpha^{1/2}}{M_\lambda^{3/2}}
\d_i\dot{\hat\chi}  \d_j\hat\chi \d_i\d_j\hat\chi
+\ldots
\bigg]
\end{align}
From the scales suppressing the interaction terms one 
reads out
the momentum and frequency 
cutoffs of the low-energy description: 
\begin{align}
\label{cutphealthy}
&\Lambda_p=\min\big\{
M_\alpha^{-1/2}M_\lambda^{3/2}~,~
M_\alpha^{3/2}M_\lambda^{-1/2}
\big\}\;,\\
\label{cutomegahealthy}
&\Lambda_\omega=\min\big\{
{M_\alpha^{1/2} M_\lambda^{1/2}}~,~M_\alpha^{-3/2}M_\lambda^{5/2}
\big\}\;.
\end{align}
If (\ref{covar21}) were the full action of the theory, it would become
inconsistent at energies / momenta above these scales
\cite{Papazoglou:2009fj}. 
Note that $\Lambda_\omega$ and $\Lambda_p$ are related by the
dispersion relation (\ref{dispchi}) and are different if the khronon
velocity differs from 1. 
We see that the scales (\ref{cutphealthy}), (\ref{cutomegahealthy}) 
are low
whenever there is a large discrepancy between 
$M_\alpha$ and $M_\lambda$.
On the other hand, in the case 
$M_\alpha\sim M_\lambda$ the cutoffs essentially coincide with $M_\alpha$,
$$\Lambda_p\sim\Lambda_\omega\sim M_\alpha\;.$$
We concentrate on this latter case in what follows.

Of course, the action (\ref{covar21}), and hence (\ref{Shealthy2}),
represents only the low-energy part of the full action
(\ref{ADMact2}). So the existence of a finite cutoff for the
low-energy theory (\ref{covar21}) does not imply any inconsistency. 
At energies/momenta larger than $M_*$ one has to take into account the
higher-derivative terms in the full action, and any conclusions drawn
from (\ref{covar21}) become invalid. By construction, the role of the
higher-derivative terms in (\ref{ADMact2}) is to modify the
power-counting rules at high energies in such a way that all the
interactions become marginal under the anisotropic scaling. 
One concludes that strong coupling is
avoided provided the scale $M_*$ of higher-derivative operators is
lower than\footnote{In particular, one can show
\cite{bps3} that 
in theories obeying the anisotropic scaling (\ref{scal})
with $z=3$ in the UV
the tree level unitarity, whose breaking
usually serves as the signal of strong coupling, is automatically
preserved provided $M_*<\Lambda_{p,\omega}$. This is essentially due to  
the peculiar kinematics of these theories that makes the unitarity
bounds milder at high energies as compared to the relativistic
case.}  $\Lambda_{p,\omega}$ \cite{bps3}.
This gives the upper bound
\be
\label{M*Mal}
M_*\lesssim M_\alpha\;.
\ee
This is a remarkable inequality: it relates the scale of quantum
gravity $M_*$ with the parameter $M_\alpha$ of the low-energy
Lagrangian which can be probed experimentally. The experimental bound
(\ref{PPNbound}) derived in Sec.~\ref{sec:healthy3} constrains
$M_\alpha$ to be lower than $10^{15}\div
10^{16}\,\mathrm{GeV}$. According to (\ref{M*Mal}) this translates
into the bound
\be
\label{M*upper}
M_*\lesssim 10^{15}\div 10^{16} \mathrm{GeV}\;.
\ee
Let us emphasize that, contrary to the
claim in \cite{Papazoglou:2009fj}, 
the choice of $M_*$  parametrically 
below $M_P$ does not introduce a fine-tuning in the model. In fact, 
having $M_*$ somewhat below $M_P$ 
is technically natural. From the point of view of the low-energy 
theory, the reason is that the cutoff for the power-law 
growing of the couplings is set by the lower scale 
$M_*$.
Thus neither $M_P$ nor $M_*$ receive large quantum corrections.

\subsection{Phenomenological considerations}
\label{sec:healthy2}

The absence of notorious pathologies in the model (\ref{ADMact2})
makes it worth to have a
closer look on its phenomenological
consequences. 
We are interested in the phenomenology of the model at energies much
lower than $M_*$. The natural language for this analysis is provided by
the St\"uckelberg formalism where at low energies the theory reduces
to GR plus the khronon field. Then all non-trivial effects of the
model are clearly interpreted as due to the presence of the khronon.
Before proceeding
we have to specify the coupling of the
khronon to the fields of the Standard Model, to which we
collectively refer as ``matter''.  

The FDiff symmetry requires that matter couples to the
khronon via geometrical objects, such as $u_\mu$, $a_\mu$,
$\KK_{\mu\nu}$, etc. 
Possible interactions fall into two classes
having qualitatively different phenomenological consequences. The
first class consists of couplings which contain derivatives of the vector
$u_\mu$; examples of this type of couplings are:
\be
\label{couple1}
a_\mu\bar\psi\gamma^\mu\psi~,~~~\KK^{\mu\nu}\bar\psi\gamma_\mu\d_\nu\psi\;,
\ee   
where $\psi$ is some fermionic matter field. 
Importantly, the combinations of the khronon field entering into the
operators (\ref{couple1}) have vanishing VEV in the flat
background. Therefore, these operators start linear in the khronon
perturbation $\chi$ and describe its derivative interaction with
matter. 
It is easy to see that in terms of 
canonically normalized fields these couplings have dimensions larger
than four\footnote{We use here the standard power-counting
  rules relevant for the low-energy physics.}
and are suppressed by the high-energy scale $M_*$. Given
that $M_*$ is large, one expects the effect of these operators to be
highly suppressed at the energies accessible to the present-day
experiments. It is still possible that couplings of the type
(\ref{couple1}) may be probed in some precision measurements. This
issue is, however, beyond the scope of the present article.

The second, more `dangerous', 
class of operators describe coupling of the matter fields
directly to the
vector $u_\mu$; examples are:
\be
\label{couple2}
u_\mu\bar\psi\gamma^\mu\psi~,~~~u^\mu
u^\nu\bar\psi\gamma_\mu\d_\nu\psi
~,~~~u^\mu u^\nu\bar\psi\d_\mu\d_\nu\psi\;.
\ee       
The crucial property of these operators 
is that they give rise to Lorentz-violating 
effects within the Standard Model as they couple matter fields to the
VEV of $u_\mu$. Such effects
and constraints on them have been extensively studied in the
literature, see
\cite{Mattingly:2005re,Jacobson:2005bg,Liberati:2009pf,Kostelecky:2008ts} 
for reviews. 
The effects of the operators (\ref{couple2}) fall into two categories: 
those that become pronounced only at
high energies, and those that persist at all energy scales. 

Effects of the first category are
associated to Lorentz-violating operators of dimensions
larger than four such as the last operator in
(\ref{couple2}). The strength of these effects is characterized by the
scale suppressing the higher-order operators which we will
collectively denote as $M_*^{(mat)}$ (one should keep in mind though
that this scale can, in general, be different for different
operators). In the model at hand it is natural to assume that
$M_*^{(mat)}$ is of the same order of magnitude as
the scale $M_*$ appearing in the gravitational
sector. However, we stress that this is
an additional assumption: in general the scales
$M_*$ and $M_*^{(mat)}$ may be different, so we prefer to keep
different notations for them. 

The experimental data constrain the scale $M_*^{(mat)}$ from below.
A rather robust bound comes from astrophysical observations
and exploits the fact that the higher-order operators lead to
modification of the dispersion relations of the matter fields, in
particular, photons, at high 
momenta\footnote{Terms with odd powers of momentum can be forbidden by
  imposing 
  the CPT invariance. In particular, this allows to avoid bounds from 
the absence of vacuum birefringence discussed in
  \cite{Jacobson:2005bg,Liberati:2009pf}.
It is also worth mentioning that modification of dispersion relation
of the matter fields at high momenta is strongly suppressed in
supersymmetric Lorentz-violating extensions of the Standard Model
\cite{GrootNibbelink:2004za,Bolokhov:2005cj}.} 
\be
\label{dispmat1}
E^2=m^2+{\bf p}^2+\eta\frac{{\bf p}^4}{M_*^{(mat)}}+\ldots\;,
\ee 
where $\eta$ is a dimensionless coefficient. This would produce a
frequency dependent delay in the arrival times of $\gamma$-rays emitted
by a distant source. The absence of such a time-lag in the signals
coming from active galactic nuclei \cite{Albert:2007qk} and
$\gamma$-ray bursts \cite{Collaborations:2009zq} yields the constraint
(assuming the coefficient $\eta$ in (\ref{dispmat1}) is of order one),
\be
\label{M*mat1}
M_*^{(mat)}\gtrsim 10^{10}\div 10^{11}\mathrm{GeV}\;.
\ee 
In the simple case when $M*^{(mat)}$ and $M_*$ are of the same order
this translates into the lower bound on $M_*$. Note that this is
several orders of magnitude below the upper limit (\ref{M*upper})
imposed by the absence of strong coupling.

It has been argued that considerably stronger constraints 
on the scale $M_*^{(mat)}$ come from the 
physics of ultra-high-energy cosmic rays (UHECR) 
\cite{Galaverni:2007tq,Maccione:2008iw,Galaverni:2008yj,Maccione:2009ju}.
These constraints are, however, less robust than (\ref{M*mat1}) as
they are sensitive to a number of assumptions about the sources of
cosmic rays and the UHECR chemical composition (e.g. the constraints
essentially 
disappear if UHECR of the highest energies consist of heavy nuclei 
\cite{Maccione:2009ju}). Thus we do not discuss these constraints in
the present paper. 

Finally, violation of Lorentz symmetry in the matter sector by
operators of dimensions 3 and four 4, such as the first and the second
operators in (\ref{couple2}), would lead to sizable effects even at low
energies\footnote{It is worth stressing that without additional assumptions
the khronon -- matter couplings generically would be non-universal,
i.e. species-dependent, implying, in particular, violation of the weak
equivalence principle.}. On the other hand, the experimental
constraints on these effects are extremely tight \cite{Kostelecky:2008ts}.
One arrives to the conclusion that breaking of Lorentz symmetry in the
matter sector at the level of dimension 3 and 4 operators must be highly
suppressed. 
For some of the couplings this may be achieved by imposing discrete
symmetries. For example, the first operator listed in (\ref{couple2})
can be forbidden by requiring CPT invariance. However, a stronger
mechanism is needed to suppress all dimension 3 and 4 operators. We
will mention one possibility
in the Discussion section.
For the moment let us just
assume that such a mechanism exists. Then to the leading
approximation the coupling of the khronon field to matter at the
lowest-derivative level must be encoded in a universal effective metric
\be
\label{geff}
\tilde g_{\mu\nu}=g_{\mu\nu}-\beta u_\mu u_\nu\;,
\ee
where $\beta$ is a dimensionless constant. Clearly, this kind of
coupling preserves Lorentz invariance of the matter sector. We
concentrate on this type of coupling
in what follows.

\subsection{Velocity-dependent forces and instantaneous interaction}
\label{sec:khronoforce}

Before engaging into a systematic analysis of the observational
constraints on the universally coupled khronon 
let us make some preliminary estimates
of the type of effects induced by the coupling 
(\ref{geff}). When $\beta$ is small
one expands the interaction terms and obtains to leading order in
$\beta$
\[
%\label{Sint}
S_{\chi-mat}=\frac{\beta}{2}\int d^4x\sqrt{-g}\;u_\mu u_\nu T^{\mu\nu}\;,
\] 
where $T^{\mu\nu}$ is the matter energy-momentum tensor. Expanding
the khronon field as usual, $\varphi=t+\chi$, one obtains the 
action for small perturbations 
including the source term
\[
%\label{khrononL}
\int d^4x \left[\frac{M_P^2}{2}
\big(\alpha (\partial_i\dot\chi)^2-(\lambda-1)(\Delta\chi)^2\big) 
-\beta \chi \partial_i T^{0i}\right]\;.
\]
From this expression we read off the khronon exchange amplitude 
between two sources with energy-momentum tensors
$T^{\mu\nu}_{(1)}$ and $T^{\mu\nu}_{(2)}$, 
\be
\label{chiexchange}
{\cal A}_{\chi} = -\frac{\beta^2}{M_P^2} T_{(1)}^{0i} \;
\frac{\partial_i \partial_j}{\Delta}\,
\frac{1}{\alpha \partial_0^2 - (\lambda-1)\Delta} \; T_{(2)}^{0j}\;.
\ee
This amplitude {\em does not} encapsulate all the new interactions due to
the presence of the khronon field. In the systematic treatment
Eq.~(\ref{chiexchange}) must be supplemented by the amplitudes
involving khronon-graviton mixing and by the contributions coming from
the modification of the metric propagator due to the Lorentz symmetry
breaking. 
This will be done in the next
section within the formalism of the parameterized post-Newtonian (PPN)
expansion. 
Here we note that the direct khronon exchange
(\ref{chiexchange})
dominates over other effects in the case
\be
\label{betalarge}
\alpha\sim |\lambda-1|\ll\beta\;. 
\ee 
The simplicity of the amplitude (\ref{chiexchange}) makes the analysis
in this case much more transparent than in the full PPN
treatment. Thus we first concentrate on the case
(\ref{betalarge}). 

Let us consider explicitly the khronon-induced
interaction for two point-like non-relativistic sources. In this case 
$T^{0i}_{(a)}=m_a v_a^i\delta({\bf x}-{\bf x}_a)$, where ${\bf x}_a$,
${\bf v}_a$, $a=1,2$ are the coordinates and velocities of the two
particles. In the non-relativistic regime one can neglect the time
derivatives in the khronon propagator. After integration over spatial
coordinates one obtains from 
(\ref{chiexchange}) the following contribution into the two-particle
Lagrangian,
\be
\label{khronoV}
L_{\chi} = \frac{m_1 m_2}{8\pi M_P^2}\; \frac{\beta^2}{\lambda-1} \;
\frac{ {\bf v_1}\cdot{\bf v_2} - 
({\bf v_1}\cdot{\bf \hat r})\; ({\bf v_2}\cdot{\bf \hat r})}{r}\;,
\ee 
where ${\bf r}={\bf x}_1-{\bf x_2}$ and ${\bf \hat r}={\bf r}/r$. 
This interaction is quite peculiar: it 
depends simultaneously on
the distance between the particles and on their velocities with
respect to the preferred frame. Note that due the velocity dependence
this term contributes non-trivially into the expressions for the
conserved energy and momentum of the system.

An interesting special case of Eq.~(\ref{khronoV}) corresponds to the
situation when 
one of the
particles (say, particle 2) is much heavier than the other 
so that its velocity is
approximately conserved (like, e.g., in the Sun -- Earth system). 
One can show that in
this case (\ref{khronoV}) reduces to 
\be
\label{khronoV2}
L_{\chi} = \frac{m_1 m_2}{8\pi M_P^2}\; \frac{\beta^2}{\lambda-1} \;
\frac{{\bf v}_2^2 -  ({\bf v}_2\cdot{\bf \hat r})^2}{r}\;,
\ee
up to a total time-derivative.
We observe 
that the dependence on the test particle velocity has dropped
out. Equation (\ref{khronoV2}) has a simple interpretation as a
direction-dependent 
contribution into the gravitational potential of the
heavy source. 
In the next section we will discuss the phenomenological constraints
on this type of contributions within the PPN framework.

An important remark is in order. 
The form of the khronon exchange amplitude
(\ref{chiexchange}) makes it manifest that the model involves
instantaneous interaction. This is due to the factor $\Delta^{-1}$ in
the khronon propagator. This factor may cancel out in some special
cases when the sources contain sufficient number of spatial
derivatives, but not in general. 
Appearance of instantaneous interactions is a common feature of
modified gravity models (more generally, gauge theories) with broken
Lorentz invariance
\cite{Dubovsky:2004sg,Gabadadze:2004iv,Dvali:2005nt,Bebronne:2008tr}. 
In a theory that does not aspire to a Lorentz invariant UV completion
this does not pose an obvious obstruction. In particular, it certainly
does not imply any problems with causality which is defined with
respect to the preferred time slicing. Instantaneous interactions
do introduce a kind of
non-locality since far-away sources may affect the immediate future in
any given local domain. However, the strength of the non-local effects
seems to decay with distance and is suppressed both by post-Newtonian
factors of $v^2$ and the (small) model parameters.  
Hence, it is not obvious if the presence of the instantaneous interaction 
leads to any significant constraints
on the model.
A more detailed investigation of
related issues (the possibility to have any direct
observational limits on instantaneous interactions, the implications
for BH physics, etc.) is left for future.

\subsection{Universal coupling and Post-Newtonian Parameterization}
\label{sec:healthy3}

In the universally coupled case the effects of the khronon field are
naturally interpreted as modification of the (universal) 
gravitational interaction
between matter particles. The size of allowed effects is constrained
by the
existing tests of GR \cite{Will:2005va}. So we now turn to the bounds
that the model has to satisfy in order to pass these tests.

The khronon sector is 
completely described by three
parameters: $(\lambda-1)$, $\alpha$ in the action (\ref{covar21}) and
$\beta$ in the effective metric (\ref{geff}). It is convenient to make
a field redefinition by trading the ``Einstein-frame'' metric
$g_{\mu\nu}$ in
favor of the effective metric $\tilde g_{\mu\nu}$ to which matter
couples minimally. 
The result of this redefinition is readily obtained if we assume
the parameters $(\lambda-1)$, $\alpha$, $\beta$ to be much
smaller than one;
we will see that this assumption is justified by the phenomenological
bounds which indeed require the above parameters to be small.  
Then to the leading order we write
\[
\sqrt{-g}\,{}^{(4)}R=\sqrt{-\tilde g}\,{}^{(4)}\tilde R-
\sqrt{-\tilde g}\left({}^{(4)}\tilde R_{\mu\nu}-\frac{1}{2}\,\tilde
g_{\mu\nu}\,{}^{(4)}\tilde R\right)
\beta\tilde u^\mu\tilde u^\nu\;.
\] 
Using the identity
\[
\int \di^4x\sqrt{-\tilde g}\,{}^{(4)}\tilde R_{\mu\nu}
\tilde u^\mu\tilde u^\nu=
\int \di^4x\sqrt{-\tilde g}\Big((\nabla_\mu\tilde u^\mu)^2-
\nabla_\mu\tilde u_\nu\nabla^\nu\tilde u^\mu\Big)
\]
one obtains the action
\be
\label{covar22}
S=-\frac{{M_P'}^2}{2}\int\di^4 x\sqrt{-g}\Big\{{}^{(4)}R
+\beta\nabla_\mu u_\nu\nabla^\nu u^\mu+\lambda'(\nabla_\mu u^\mu)^2
+\alpha u^\mu u^\nu\nabla_\mu u^\rho\nabla_\nu u_\rho\Big\}\;,
\ee
where
\be
\label{paramnew}
{M_P'}^2=M_P^2\left(1+\frac{\beta}{2}\right)~,~~~\lambda'=\lambda-1-\beta
\ee
and we have omitted tildes over the new variables (we will always work
with the redefined metric in the rest of this section
so this will not
lead to confusion). Note that the appearance of the
new parameter $\beta$ in
the khronon action (\ref{covar22}) can be traced back to the presence
of the parameter $\xi$ in front of the 3-dimensional
scalar curvature in the unitary-gauge potential, see (\ref{potadd}),
(\ref{potential}).  
In the pure gravity theory this parameter does not have physical
meaning as it can be eliminated, say, by a rescaling of the time
coordinate, cf. the discussion after Eq.~(\ref{potential}).
This is no longer true in the presence of matter: as we
are going to see physical observables depend on the value of $\beta$.

As already mentioned before, the action (\ref{covar22}) has the same
form as the action of the Einstein-aether theory \cite{Jacobson:2000xp}
which has been extensively studied as a phenomenological model for
violation of Lorentz invariance \cite{Jacobson:2008aj}. The difference
of our model from Einstein-aether is that in our case the
aether vector $u_\mu$ is by construction hypersurface-orthogonal. As a
consequence, in our case aether propagates a single longitudinal
degree of freedom (khronon), while in general there are additional
transverse modes. This implies that the results about Einstein-aether theory
that are insensitive to the presence of transverse modes are also
valid for our model\footnote{The relation with Einstein-aether can
  also be used to derive the result of metric redefinition
  (\ref{geff}) in the general case when the parameters $(\lambda-1)$,
  $\alpha$, $\beta$ are not small. In Einstein-aether the substitution 
 (\ref{geff}) leads to the change of
the coefficients in the aether Lagrangian; the corresponding formulas
have been worked out in \cite{Foster:2005ec}. The most general Einstein-aether
action contains an additional term
$\nabla_\mu u_\nu \nabla^\mu u^\nu$ compared to (\ref{covar22}). 
In our case when the
aether vector $u_\mu$ is hypersurface-orthogonal the four terms in
the aether action are not independent 
\cite{Jacobson:2010mx} implying that 
the term $\nabla_\mu u_\nu \nabla^\mu u^\nu$
can be eliminated. In this way 
one arrives at the action (\ref{covar22}) with
\[
{M_P'}^2=\frac{M_P^2}{\sqrt{1-\beta}}~,~~~\lambda'=(1-\beta)(\lambda-1)-\beta\;,
\]
which coincides with (\ref{paramnew}) for $\beta\ll 1$.}. 

The first set of constraints obtained in this way is related to
the fact that the velocities of the modes propagated by the action
(\ref{covar22}) are in general different from 1, the maximal velocity
of matter. Using the formulas for the
Einstein-aether theory \cite{Jacobson:2008aj} we obtain to first
non-trivial order in parameters the velocity of 
helicity-2 modes (graviton):
\bseq
\label{velocities}
\be
\label{cg}
c^2_g=1+\beta\;,
\ee
and that of the helicity-0 mode (khronon):
\be
\label{cchi}
c^2_\chi=\frac{\lambda'+\beta}{\alpha}\;.
\ee
\eseq
Note that (\ref{cchi}) coincides
with the velocity entering the dispersion relation (\ref{dispchi}). If
the velocities of graviton or khronon are smaller than 1 
relativistic matter particles
will quickly loose their energy via vacuum Cherenkov radiation 
\cite{Elliott:2005va}. This is strongly contrained by the existence 
of high-energy cosmic rays. Thus we conclude that the graviton and
khronon velocities must be larger or equal to 1 which yields the
bounds
\[
%\label{cherb}
\beta\geq 0~,~~~\frac{\lambda'+\beta}{\alpha}\geq 1\;.
\]

Another constraint is obtained from the comparison of the
gravitational constants appearing in the Newton law and the Friedmann 
equation
governing the cosmological expansion. Again, the transverse aether
modes do not play any role in these considerations, so we can
directly apply the results from the Einstein-aether theory to our
case. The Newton
constant, which is defined as the coefficient in the Newton law for the
gravitational force between two static masses, is related to the
parameters appearing in the action (\ref{covar22}) as follows 
\cite{Jacobson:2008aj}:
\be
\label{GN}
G_N=\frac{1}{8\pi {M_P'}^2(1-\alpha/2)}\;.
\ee
On the other hand, the cosmological expansion in the 
Einstein-aether theory is described
by the standard Friedmann equation,
\be
H^2=\frac{8\pi}{3}G_{cosm}\,\rho\;,
\ee
where $H$ is the Hubble rate, $\rho$ -- the energy density of the
Universe, but with a different proportionality
coefficient\footnote{The expressions (\ref{GN}), (\ref{Gcosm}) for the
  case $\beta=0$ were
  derived directly from the action (\ref{ADMact2}) in \cite{bps2}.}
\cite{Jacobson:2008aj},
\be
\label{Gcosm}
G_{cosm}=\frac{1}{8\pi {M_P'}^2(1+3\lambda'/2+\beta/2)}\;.
\ee
The discrepancy between $G_N$ and $G_{cosm}$ is constrained by Big
Bang nucleosynthesis \cite{Carroll:2004ai}, 
\[
%\label{BBN}
\left|\frac{G_{cosm}}{G_N}-1\right|\leq 0.13\;.
\]
Barring accidental cancellations, this yields an 
order-of-magnitude bound on the parameters of the model,
\[
%\label{BBNbound}
\alpha,~\beta,~\lambda'\lesssim 0.1\;.
\] 
 
Stringent limits on any alternative theory of gravity come from the
observational constraints on the parameters of the parameterized
post-Newtonian (PPN) formalism. Remarkably, in the Einstein-aether
theory all PPN parameters except two are the same as
in GR \cite{Foster:2005dk}. The non-trivial parameters are called 
$\alpha_1^{PPN}$, $\alpha_2^{PPN}$ and describe preferred frame
effects related to breaking of Lorentz symmetry. We now argue that in
the khrono-metric theory (\ref{covar22}) all PPN parameters except
$\alpha_1^{PPN}$, $\alpha_2^{PPN}$ are the same as in Einstein-aether.
Indeed, besides $\alpha_1^{PPN}$, $\alpha_2^{PPN}$
and the parameters which vanish automatically in any
theory described by a Lagrangian, there are three more PPN parameters:
$\beta^{PPN}$, $\gamma^{PPN}$ and $\xi^{PPN}$. The key point is that
these three are determined from spherically symmetric
solutions which are identical in the khrono-metric and
Einstein-aether theories, see Appendix~\ref{app:F}. 
Thus in the khrono-metric model at hand these 
parameters have the same values as in GR,
\[
%\label{bgxi}
\beta^{PPN}=\gamma^{PPN}=1~,~~~\xi^{PPN}=0\;.
\]

One cannot use the results for Einstein-aether in the case of the
parameters $\alpha_1^{PPN}$, $\alpha_2^{PPN}$: they
describe effects related to the motion of the source with respect
to the preferred frame and are contaminated in Einstein-aether 
by the contributions of
the transverse aether modes. These parameters can be defined as the
coefficients in the linearized metric
produced by a point source of mass $m$
{\em in its rest frame} \cite{Will:1972zz}:
\bseq
\label{hPPN}
\begin{align}
\label{hPPN1}
&h_{00}=-2G_N\frac{m}{r}\left(1-\frac{(\alpha_1^{PPN}-\alpha_2^{PPN})v^2}{2}
-\frac{\alpha_2^{PPN}}{2}\frac{(x^iv^i)^2}{r^2}\right)\;\\
&h_{0i}=\frac{\alpha_1^{PPN}}{2} G_N\frac{m}{r}v^i\;,\\
&h_{ij}=-2G_N\frac{m}{r}\delta_{ij}\;,
\end{align}
\eseq 
where $r$ is the distance from the source and $v^i$ is the velocity of
the source with respect to the preferred frame. 
Note that the contribution proportional to $\alpha_2^{PPN}$ has the
form of the direction-dependent gravitational potential encountered in
Sec.~\ref{sec:khronoforce}. 
The current Solar
system limits on these parameters are \cite{Will:2005va}:
\[
%\label{PPNlimits}
|\alpha_1^{PPN}|\lesssim 10^{-4}~,~~~
|\alpha_2^{PPN}|\lesssim 10^{-7}\;.
\]
Derivation of $\alpha_1^{PPN}$, $\alpha_2^{PPN}$ for the model
(\ref{covar22}) in the 
case $\alpha,\beta,\lambda'\ll 1$
is given in Appendix~\ref{app:D}; the result is
\be
\label{PPNpar}
\alpha_1^{PPN}=-4(\alpha-2\beta)~,~~~~~
\alpha_2^{PPN}=\frac{(\alpha-2\beta)(\alpha-\lambda'-3\beta)}
{2(\lambda'+\beta)}\;. 
\ee
Note that both parameters vanish if $\alpha-2\beta=0$. Another
interesting case is $\beta=0$,
$\lambda'=\alpha$; this corresponds to the situation when 
the velocities of all modes in
the theory are equal to 1, see Eqs.~(\ref{velocities}). In this case the
parameter $\alpha_2^{PPN}$ which is most tightly constrained vanishes.
Barring these special cases and assuming $\alpha$,
$\beta$, $\lambda'$ to be of the same order we obtain the bound
\be
\label{PPNbound}
\alpha~,~\beta~,~\lambda'~\lesssim ~10^{-7}\div 10^{-6}\;.
\ee
To the best of our knowledge, this is the strongest constraint on the
parameters of the model that can be obtained at present.

As discussed in Sec.~\ref{sec:healthy1}, the bound (\ref{PPNbound})
combined with the requirement that the theory is weakly coupled implies
the upper limit (\ref{M*upper}) on the scale suppressing
higher-order operators in the gravitational action. 

\section{Summary and discussion}
\label{sec:concl}

In this paper we have investigated the self-consistency issues 
related to the scalar graviton modes  
in Ho\v rava's approach to quantum
gravity. We have considered three models of non-relativistic gravity 
differing by 
the symmetry group
and the requirement or not of the
projectability condition. 
In our study we extensively used 
the St\"uckelberg formalism which makes the
scalar modes explicit by encoding them into the khronon field: a
scalar field with time-dependent VEV. This facilitates a lot the
analysis of the scalar mode dynamics. 

In two of the considered models (the projectable version of the original
proposal \cite{Horava:2009uw} and a possible extension based on a
smaller symmetry group),  
the scalar modes were found to exhibit
pathological behavior, such as instability and strong coupling. These
pathologies undermine the perturbative analysis. In particular, they
invalidate the naive power-counting argument for renormalizability
of these models.

Several qualitative lessons can be extracted from these studies. While
it is relatively easy to make the scalar modes well-behaved in the UV, this
is much harder to achieve in the IR. The scalar gravitons tend to
develop gradient instabilities which can be suppressed only by pushing
the model parameters to extreme values. This, in turn, introduces
strong coupling. Thus the primary goal in constructing a consistent
non-relativistic gravity model is to stabilize the scalar modes.

The analysis of Sec.~\ref{sec:ghc} teaches us that reducing
the symmetry to a smaller group than the FDiffs does not
give any advantage in achieving this goal. Though this approach allows
to improve the behavior of the scalar graviton of the projectable Ho\v
rava's model, it introduces yet other scalar modes which bring the
pathologies back.
Even relaxing the symmetry to the RFDiffs, that have as
many local generators as the FDiffs, turns out to have
the quite dramatic consequence of
allowing for new operators that lead to tachyonic ghosts and therefore fast
instabilities.
Even if
not present at tree level these operators would be generated by
quantum effects. From this we conclude that 
the time-reparameterization symmetry contained in the FDiffs
plays the quite important role of preventing the appearance of
tachyonic ghosts
in the model.

Remarkably, the scalar mode is stable and also free from other
pathologies 
in the third model which we analyzed. 
The latter is a natural extension of
the original non-projectable Ho\v rava model obtained by including into
the action all terms compatible with invariance under FDiffs and
renormalizability by power counting. For appropriate choice of
parameters the unique propagating scalar mode possesses
stable dispersion relation in the entire range of spatial
momenta. Moreover, the dispersion relation has nice properties both at
low and high momenta. In the first case it is linear, 
$\omega^2\propto p^2$, implying that the mode remains stable in any
sufficiently smooth background\footnote{More precisely, in any
  background sufficiently close to Minkowski space-time equipped with
  the foliation by surfaces $t=const$.}. On the other hand, the
asymptotic form
$\omega^2\propto p^6$ of the dispersion at high momenta is compatible
with the anisotropic scaling postulated in the UV. Therefore, the
presence of this mode does not spoil the power-counting arguments
which strongly suggest that the model is renormalizable. This implies,
in particular, that strong coupling is avoided in the model by
construction provided the coupling constants are chosen small enough.

Encouraged by these results we studied  some phenomenological aspects
of the healthy model.
Making use of the St\"uckelberg formalism we have demonstrated
that at low energies the theory reduces to GR interacting with an
additional scalar field --- the khronon. The time-dependent VEV of the
khronon field breaks
Lorentz invariance down to arbitrarily low energies. We observed that the
structure of the low-energy limit of the model is similar to that of the
Einstein-aether theory \cite{Jacobson:2000xp,Jacobson:2008aj}, even
though
the two theories are not exactly equivalent. The difference
is due to transverse modes present in the Einstein-aether theory
and absent in the healthy model of this paper. The transverse modes do
not affect the form of homogeneous isotropic
cosmological solutions and spherically symmetric solutions. This
allowed us to directly apply to our case the bounds on Einstein-aether
theory coming from the expansion history of the Universe, as well as
to conclude that all but two PPN parameters in the healthy model
coincide with those of GR. We have calculated the values of the
remaining two PPN parameters, $\alpha_1^{PPN}$, $\alpha_2^{PPN}$,
which characterize preferred frame effects. Current observational
bounds on $\alpha_1^{PPN}$, $\alpha_2^{PPN}$ significantly constrain
the parameter space of the healthy model (but do not rule it
out). Combined with the requirement that the theory is weakly coupled
these bounds translate
into the upper bound on the scale of
quantum gravity $M_*\lesssim 10^{15}\div 10^{16}\,\mathrm{GeV}$. 
Within the simplifying assumption that $M_*$ coincides with the scale
of Lorentz symmetry breaking in the matter sector,
astroparticle data constrain it from below: 
$M_*\gtrsim 10^{10}\div 10^{11}\,\mathrm{GeV}$.
[We stress though that this lower bound must be taken cautiously: its
validity depends on the details of the matter sector.]
Thus the net result of our
phenomenological study is that the healthy model can be tested
experimentally using existing techniques, which we believe makes this
model quite
attractive.  

The above results suggest that the healthy model of this paper can
serve as a starting point for construction of a viable renormalizable
theory of gravity. Admittedly, a lot of open issues remain. Let us
discuss some of them. 

The renormalizability of the model is yet to be demonstrated beyond
the naive power counting. In principle, this amounts to an explicit
analysis of the loop corrections to the action of the form
(\ref{ADMact2}), (\ref{potadd}). In practice though, this may turn out
to be a formidable task, given the large (of order 100) number of
allowed terms in the Lagrangian and the complications related to the
gauge symmetry of the action under FDiffs. 
A particularly subtle issue
is a proper treatment of the time-reparameterization invariance. For
consistency of the model this symmetry must be free of 
anomalies.
Otherwise the symmetry would be reduced 
to that of the model of Sec.~\ref{sec:ghc}, with the corresponding 
re-appearance of pathologies. An interesting development along these
lines is 
the renormalization of the energy-momentum 
tensor of test fields in curved backgrounds in Ho\v rava-type
theories 
\cite{Giribet:2010th}. It is shown that, in contrast to the
relativistic case, this does not require counterterms with more than
two time derivatives.

Even if the model proves to be renormalizable, there will be a
question if it is UV-complete or not. In other words, if it possesses
a weakly coupled UV fixed point. An answer to this question requires the
study of the renormalization group (RG) flow of the theory. Let us
mention in this connection interesting recent results \cite{Iengo:2010xg} 
about quantum
electrodynamics (QED) in five space-time dimensions.
It is demonstrated that 5d QED
can be UV completed within Lorentz-violating framework by adding
operators with higher spatial derivatives. The resulting theory  is
weakly coupled at all energies, possesses a weakly coupled UV fixed
point with anisotropic scaling exponent $z=2$ 
and flows to the usual 5d QED in IR (though, in general, with
different velocities of photons and electrons).    

It will be interesting to assess the quantum properties of the healthy
model
beyond perturbation theory. A possible approach to this difficult
problem would be to use the canonical formalism.
It is worth noting that inclusion of the
terms with derivatives of the lapse $N$ into the potential
(\ref{potadd}) significantly improves the canonical structure of 
the theory compared to the original non-projectable 
version of the
proposal. Indeed, due to these terms the lapse is no longer a Lagrange
multiplier and the analysis \cite{Li:2009bg,Henneaux:2009zb} 
unveiling the pathological structure of the constraints
in the original Ho\v rava model does not
apply. Instead, in the model of this paper the Hamiltonian constraint
obtained as the variation of the action with respect to the lapse,
${\cal H}\equiv\frac{\delta S}{\delta N}=0$, has non-vanishing Poisson
bracket with the equation $\pi_N=0$, where $\pi_N$ is the canonically
conjugate momentum for $N$. In other words, these two constraints form
a second class pair and must be used to eliminate the variables $N$,
$\pi_N$ from the phase space \cite{bps2,Kluson:2010xx}. 
Presence of second class constraints is an interesting feature of the
model and its implications both  for classical and quantum dynamics of
the theory deserve a detailed study.

The non-perturbative dynamics of the model can be also addressed
semiclassically by developing perturbative expansion in the
backgrounds of classical non-linear solutions. Clearly, the first step
here is
to find explicitly such solutions. Let us point out in this
context the existence 
of cosmological solutions
which are free from the initial
singularities\footnote{These solutions were
  obtained in the context of the original Ho\v rava's proposal which
  suffer from the pathological behavior of the extra mode. However,
  the terms which were added to the original Lagrangian to obtain the
  healthy model vanish on spatially homogeneous configurations (they
  contain spatial gradients of the lapse). Thus the cosmological
  solutions of the original Ho\v rava's model are also solutions of the
  healthy model of the present paper.}
\cite{Calcagni:2009ar,Kiritsis:2009sh,Brandenberger:2009yt}. It will
be interesting to understand if these solutions are stable. 
Another important issue is the structure of black
hole solutions in the healthy model. Given that the high frequency
modes in this model propagate with arbitrarily high velocities
(moreover, as discussed in Sec.~\ref{sec:khronoforce}, the theory
involves instantaneous interactions), it is
a priori unclear if the notion of black hole as a region of space
surrounded by an event horizon makes sense. On the
other hand, there certainly must exist some low-energy notion of horizon
defined using low-frequency modes which have finite velocity. It will
be interesting to understand the physical meaning of black hole
entropy associated to this low-energy horizon. The situation is even
more intriguing because generically, due to the violation of Lorentz
invariance, different low-energy species will propagate with different
velocities thus giving rise to a number of nested horizons.  
As discussed in \cite{Dubovsky:2006vk,Eling:2007qd} 
one would expect this to lead to break down of black hole
thermodynamics, which on general grounds implies violation of
unitarity of the underlying theory. Thus it is vital for the proposal
to understand if and how this paradox is resolved.
It is worth mentioning that some
spherically symmetric solutions of the healthy 
model have been reported in
\cite{Kiritsis:2009vz}.

From the phenomenological perspective the major challenge for 
the non-relativistic gravity framework is a mechanism for emergence of
Lorentz invariance in the matter sector at low energies.
As discussed above this must happen
with very high accuracy to satisfy existing experimental bounds. This
poses a severe fine-tuning problem which is aggravated by the fact
that Lorentz-violating parameters, such as differences of velocities of
various matter species, run with the energy scale
\cite{Collins:2004bp,Iengo:2009ix}. A promising mechanism to avoid
fine-tuning is related to supersymmetry. Indeed, it has been argued
\cite{GrootNibbelink:2004za,Bolokhov:2005cj} that within
Lorentz-violating extensions of the Minimal Supersymmetric Standard
Model (MSSM) it is impossible to write any Lorentz-violating operator
of dimension less than 5. Dimension 5 Lorentz-breaking operators can
be further forbidden by imposing discrete symmetries, e.g. CPT; 
in this case
Lorentz breaking starts at dimension 6. In other words, given
supersymmetry, Lorentz invariance emerges as an accidental low-energy
symmetry. Eventual supersymmetry breaking gives rise to
Lorentz-violating effects, but these are suppressed by the small ratio
of the soft supersymmetry breaking masses $m_{soft}$ to the scale
$M_*$ of the higher-order Lorentz-violating operators.
For example, the coefficients of dimension-4 Lorentz-violating operators
generated from the operators of dimension 6 upon supersymmetry
breaking are of order 
$(m_{soft}/M_*)^2$. These are
comfortably 
within the experimental bounds for $m_{soft}\sim 1$ TeV, 
$M_*\sim 10^{15}$ GeV. Interestingly, supersymmetry also suppresses
the contributions with higher powers of momentum in the dispersion
relations of matter particles
\cite{GrootNibbelink:2004za,Bolokhov:2005cj} which weakens
significantly the lower limits on $M_*$ coming from astrophysical
observations. Needless to say, realization of this scenario for the
emergence of Lorentz invariance requires supersymmetrizing the
non-relativistic gravitational action (\ref{ADMact2}),
(\ref{potadd}). The first step in this direction would be to
supersymmetrize the low-energy limit of this action,
Eq.~(\ref{covar22}). This essentially amounts to constructing a 
supersymmetric
action for the khronon sector as the gravitational part coincides
with GR and can be made supersymmetric in the standard way. 

The present work makes only first steps
in the study of observational consequences of the healthy model. Many
topics which have not been touched in this paper deserve a thorough
investigation. These include implications of the model for emission
and propagation of gravitational waves, dynamics of binary pulsars,
spectrum of CMB perturbations and structure formation. Interesting
results about the dynamics of the cosmological perturbations in the
healthy model were reported recently in
\cite{Kobayashi:2010eh,ArmendarizPicon:2010rs}. 

We plan to return to some of the above questions in future.\\

{\it Note added.} When this paper was in preparation the article
\cite{Kimpton:2010xi} appeared which also uses the St\" uckelberg formalism
to study the properties of the healthy model. Where there is an
overlap, our results agree. However, Ref.~\cite{Kimpton:2010xi} pushes
the analysis using the decoupling limit to the regime of high energies
(higher than $M_*$) and argues that the terms with the higher spatial 
derivatives do not resolve strong coupling
of the scalar mode. This seems to contradict the conclusions of the
present paper. In fact, pushing the decoupling limit to high energies,
as done in \cite{Kimpton:2010xi}, corresponds to going beyond its range
of applicability. 
Indeed, the limit considered in \cite{Kimpton:2010xi}
corresponds to throwing away all the higher-derivative mixings between
the khronon field and the metric. On the other hand, as admitted
in the published version of the \cite{Kimpton:2010xi}, these terms are
crucial to ensure the correct UV behavior $\omega^2\propto p^6$ of the
dispersion relation of the scalar mode, required by the anisotropic
scaling with $z=3$ and satisfied in the full theory, see
Eq.~(\ref{dispgood}). Instead, for the decoupling limit considered in 
\cite{Kimpton:2010xi} the UV asymptotic of the dispersion relation for the
khronon field is $\omega^2=const$. Thus the loop integrals over
spatial momenta instead of being suppressed become even more divergent
than in the relativistic case. 
Clearly, this only represents a failure of the 
limit considered in \cite{Kimpton:2010xi} and
does not imply the inconsistency of the 
theory.  
%In other words, at high energies
% there is no regime in which the mixing between the 
%khronon and the metric can be neglected. Hence,  strictly speaking 
%there is no `decoupling limit'. Therefore,
% studying the khronon mode alone with 
%the metric perturbation frozen, as done in \cite{Kimpton:2010xi}, does not 
%capture the actual dynamics of the model. 

%%%%%%%%%%%%%%%%%%%%%%%%%%%%%%%%%%%%%%%%%%%%%%%%%%%%%%%%%%%%%%
\paragraph*{Acknowledgments}
%%%%%%%%%%%%%%%%%%%%%%%%%%%%%%%%%%%%%%%%%%%%%%%%%%%%%%%%%%%%%%

We thank Gia Dvali, Sergei Dubovsky, Gregory Gabadadze, Petr Ho\v
rava, Ted Jacobson,
Elias Kiritsis,
Kazuya Koyama, Riccardo Rattazzi,
Michele Redi, Thomas Sotiriou, 
Alessandro Strumia, Alex Vikman and Giovanni Villadoro
for useful discussions. We are indebted to Valery Rubakov for valuable
comments on the manuscript.
This work was supported in part by the Swiss Science Foundation
(D.B.), the Tomalla Foundation (S.S.), RFBR grants 08-02-00768-a and
09-01-12179 (S.S.)
and the Grant of the President of Russian Federation
NS-1616.2008.2~(S.S.). 

\appendix

\section{Proof of the no ghost theorem}
\label{app:A}

In this appendix we prove the proposition formulated at the end of
Sec.~\ref{sec:stuck}. 
For simplicity, let us first consider the case when 
the Lagrangian depends only on the normal vector
$u^\mu$ and its first derivatives. These derivatives enter into the
Lagrangian through the extrinsic curvature $\KK_{\mu\nu}$ and the
acceleration vector $a^\mu$,
\[
%\label{LKK}
{\cal L}={\cal L}(u^\mu, {\cal K}_{\mu\nu}, a^\mu)\;.
\] 
This case covers all terms in the general Ho\v rava-type Lagrangian
(\ref{ADMact2}) 
except those involving spatial derivatives of the 3-dimensional
 tensor $R_{ij}$ or of the acceleration $a_i$. 

We start
by expanding $u_\mu$ up to quadratic order,
\[
%\label{second_def}
u_\mu=\bar u_\mu+\delta^{(1)}u_\mu +\frac{1}{2}\delta^{(2)}u_\mu\;,
\]
where
\begin{align}
\label{u1}
&\delta^{(1)}u_\mu=\frac{1}{\sqrt{\bar X}}\bar P^\nu_\mu\d_\nu\chi\;,\\
\label{u2}
&\delta^{(2)}u_\mu=-\frac{1}{\bar X}
\bar u_\mu (\bar P^{\nu\lambda}\d_\lambda\chi)^2
-\frac{2}{\bar X}\bar P^\lambda_\mu\d_\lambda\chi\;
\bar u^\nu\d_\nu\chi\;.
\end{align}
Here bar refers to the background values of the
fields. The crucial observation is that in the background
(\ref{unitback}) the first variation (\ref{u1}) of the normal vector
does not contain time derivatives of $\chi$, while the second
variation (\ref{u2}) contains only first time derivative. The next
step is to
consider variations of the extrinsic curvature and acceleration
\begin{align}
&\delta^{(1)} \KK_{\mu\nu}=\delta^{(1)}P^\lambda_\mu\;\nabla_\lambda\bar
u_\nu +\bar P^\lambda_\mu\;\nabla_\lambda \delta^{(1)}u_\nu\;,\notag\\
&\delta^{(1)}a_\mu=\delta^{(1)}u^\lambda\;\nabla_\lambda\bar u_\mu
+\bar u^\lambda\;\nabla_\lambda\delta^{(1)}u_\mu\;,\notag\\
&\delta^{(2)}\KK_{\mu\nu}=2\delta^{(1)}P^\lambda_\mu\;\nabla_\lambda\delta^{(1)}u_\nu
+\delta^{(2)}P^\lambda_\mu\;\nabla_\lambda\bar u_\nu
+\bar P^\lambda_\mu\;\nabla_\lambda\delta^{(2)}u_\nu\;,\notag\\
&\delta^{(2)}a_\mu=2\delta^{(1)}u^\lambda\;\nabla_\lambda\delta^{(1)}u_\mu
+\delta^{(2)}u^\lambda\;\nabla_\lambda\bar u_\mu
+\bar u^\lambda\;\nabla_\lambda\delta^{(2)}u_\mu\;.\notag
\end{align}
One observes that in the background (\ref{unitback})
$\delta^{(1)}\KK_{\mu\nu}$ does not contain time derivatives of
$\chi$;
$\delta^{(1)}a_{\mu}$, $\delta^{(2)}\KK_{\mu\nu}$ contain one time
derivative, and $\delta^{(2)}a_{\mu}$ -- two time derivatives. 
The quadratic Lagrangian consists of two types of terms: the terms
containing the first variations of the fields squared, and the
terms linear in the second variations. For example, writing down
explicitly the terms with the variation of the acceleration we obtain
\[
{\cal L}^{(2)}=\frac{1}{2}
\overline{\frac{\d^2{\cal L}}{\d a_\mu\d a_\nu}}\;
\delta^{(1)} a_\mu\;\delta^{(1)} a_\nu
+\frac{1}{2}\overline{\frac{\d{\cal L}}{\d a_\mu}}\;
\delta^{(2)} a_\mu+\ldots\;.
\]
Clearly, both these terms contain at most two time derivatives of
$\chi$. In the case when $\bar\varphi$ does not satisfy equations of
motion one has to consider also 
the linear variation of the Lagrangian ${\cal L}^{(1)}$. 
Using the same reasoning as before one concludes that
${\cal L}^{(1)}$ contains at most first time derivative of $\chi$.
This completes the proof.

It is straightforward to generalize the
above proof to include the dependence of the
Lagrangian on higher spatial derivatives. In covariant
  language a spatial derivative translates into the operator 
$P\circ\nabla\circ P$, which is purely spatial (i.e. does not
introduce further time derivatives). For example, the object $\nabla_i a_j$
takes the covariant form 
$P_\mu^\lambda\nabla_\lambda(P^\rho_\nu a_\rho)$.
Using the same reasoning as before one can show that the first
(second) variation of this type of objects contain at most one (two)
time derivatives of $\chi$.

Note that the above proof uses in an essential way the invariance of
the action under the transformations (\ref{phirepar}): this forces
the Lagrangian to depend on the khronon field only via
$u_\mu$ and its derivatives. Recall that
this invariance stems from the invariance of the original action
(\ref{ADMact2}) under FDiffs.
As a consequence the proof also goes through for 
the projectable model
(\ref{ADMact1}) which obeys this symmetry. 
Indeed, as pointed out in Sec.~\ref{sec:theories}, the projectable
model is recovered from the theory (\ref{ADMact2}), (\ref{potadd}) by
taking the limit $\alpha\to\infty$. Thus the order of time derivatives
in the covariant equations for these two theories coincide. 

On the
other hand, the proof does not apply to models 
with reduced symmetry, such as the theory
(\ref{ADMact3}), because in these cases the Lagrangian contains
additional dependence on the $\varphi$ gradients.
We will see in
Sec.~\ref{sec:ghc} that this makes the equation of motion for
the khronon fourth-order in time; as one can anticipate, this will lead to
certain pathologies of the theory.

\section{Two faces of the tachyonic ghost}
\label{app:B}

The purpose of this appendix is to clarify the following puzzle. 
In Sec.~\ref{sec:ghc} we have shown using the St\"uckelberg formalism that 
the RFDiff model includes in its spectrum a tachyonic ghost. On the
other hand, in the unitary gauge the action (\ref{ADMact3}) of the
model contains only two time derivatives and according to the standard
lore one would not expect any ghosts in this picture. This seems to
contradict the results of the St\"uckelberg analysis. We are going to
show that the contradiction is removed when one properly formulates
the physical questions to assess the effects of the tachyonic ghost.

We start by writing the unitary gauge 
Lagrangian for the sector of scalar
perturbations 
around Minkowski
background. Substituting the decomposition (\ref{dec}) in the action
(\ref{ADMact3}) and integrating out 
the non-dynamical fields $B$ and $E$ one obtains 
\be
\label{L2ghc}
\begin{split}
{\cal L}^{(2)}_{III}&=\frac{M_P^2}{2}\bigg[
\frac{4M_P^2}{M_\lambda^2}\dot\psi^2-2\psi\Delta\psi
+4\phi\Delta\psi\bigg]+\frac{M^2_\alpha}{2}(\d_i\phi)^2
+\frac{M^2_{\lambda_1}}{2}\dot\phi^2+M^4\phi^2\\
&-\frac{M_P^2}{2}\bigg[\frac{f_1}{M_*^2}(\Delta\psi)^2+
\frac{2f_2}{M_*^2}\Delta\phi \Delta\psi+
\frac{f_3}{M_*^2}(\Delta\phi)^2+
\frac{g_1}{M_*^4}\psi\Delta^3\psi+
\frac{2g_2}{M_*^4}\phi\Delta^3\psi+
\frac{g_3}{M_*^4}\phi\Delta^3\phi\bigg]\;,
\end{split}
\ee
where $M_\lambda$, $M_{\lambda_1}$, $M_\alpha$ are defined in
(\ref{Mscales})
and $M^4$ is the coefficient appearing in the expansion of the
potential $V(N)$ in (\ref{potIII}) to quadratic order in $\phi$; these
parameters have the same meaning as in the St\"uckelberg action
(\ref{SchiIII}). In deriving (\ref{L2ghc}) we assumed for simplicity
$M_\lambda\ll M_P$ (this is the case relevant for comparison with the
St\"uckelberg analysis).
Finally, the 
constants $f_n$, $g_n$ are related to the coefficients of
the higher-derivative terms in the potential (\ref{potIII}). The
Lagrangian (\ref{L2ghc}) 
clearly describes two propagating degrees of freedom which
matches with the St\"uckelberg analysis of Sec.~\ref{sec:ghc}. What
seems to be different is that the kinetic energy of both modes
can be made positive by choosing
$M_\lambda,~M_{\lambda_1}>0$, while in the khronon language one of the
modes is a ghost. 
This said, let us proceed and find the dispersion relation for the
modes. Neglecting the higher-derivative terms we
obtain,
\be
\label{QFdisp}
\begin{split}
\omega^2=&-\bigg[\frac{M^4}{M^2_{\lambda_1}}
+\bigg(\frac{M^2_\alpha}{2M^2_{\lambda_1}}
+\frac{M^2_\lambda}{4M^2_P}\bigg){\bf p}^2\bigg]\\
&\pm\sqrt{\bigg[\frac{M^4}{M^2_{\lambda_1}}
+\bigg(\frac{M^2_\alpha}{2M^2_{\lambda_1}}
+\frac{M^2_\lambda}{4M^2_P}\bigg){\bf p}^2\bigg]^2
+\frac{M_\lambda^2{\bf p}^4}{M^2_{\lambda_1}}
-\frac{M^4M^2_\lambda{\bf p}^2}{M_P^2M^2_{\lambda_1}}}\,.
\end{split}
\ee 
Clearly one of the modes exhibits gradient instability. In the decoupling
limit $M_P\to\infty$ the dispersion relations (\ref{QFdisp}) coincide
with the expressions (\ref{dispIII}) of Sec.~\ref{sec:ghc}. Thus at
the level of the dispersion relations 
the unitary gauge and the St\"uckelberg descriptions
match.

Let us now ask if there is any physical setup where one could 
distinguish between a
mode with gradient instability (simple tachyon) and a mode which
besides the gradient instability also has a negative kinetic
term (tachyonic ghost). The standard definition of the ghost as ``a
field whose Hamiltonian is not positive definite'' is not very useful:
the Hamiltonian is not sign-definite in both cases. Moreover, the two
cases can be related by a canonical transformation. To illustrate this
point consider a toy Lagrangian representing a single Fourier mode 
of a tachyonic ghost,
\be
\label{toyghost}
L_{ghost}=-\frac{\dot\eta^2}{2}-\frac{p^2\eta^2}{2}\;.
\ee   
The corresponding Hamiltonian reads
\[
H_{ghost}=-\frac{\pi^2}{2}+\frac{p^2\eta^2}{2}\;,
\]
where $\pi$ is the canonically conjugate momentum for $\eta$. The
canonical transformation
\[
\tilde\pi=p\eta~,~~~\tilde\eta=\pi/p
\]
casts this into the Hamiltonian of a simple tachyon,
\[
H_{tachyon}=\frac{\tilde\pi^2}{2}-\frac{p^2\tilde\eta^2}{2}\;.
\]

One may still try to distinguish if a mode is ghost or not by the sign
of the residue at the pole in the one-particle exchange amplitude
considered as function of $\omega^2$; this must be negative for the
mode to qualify as a ghost. However, 
in the case of the tachyonic ghost the sign 
depends on the type of the coupling to the source used to
define the amplitude. Taking again 
the toy model (\ref{toyghost}) as an
example consider two couplings:
\[
L_{source}^{(1)}=\eta \Sigma_1~~~\mathrm{and}~~~
L_{source}^{(2)}=\dot\eta \Sigma_2\;.
\]
In the first case the one-particle exchange amplitude reads
\[
{\cal A}_{\Sigma_1}\propto \Sigma_1\; \frac{-1}{\omega^2+p^2}\;
\Sigma_1
\] 
and the residue is negative. However, in the second case
\[
{\cal A}_{\Sigma_2}\propto \Sigma_2\; \frac{-\omega^2}{\omega^2+p^2}\;
\Sigma_2
\] 
and the residue becomes positive: $-\omega^2=p^2>0$. Note that this
ambiguity is related to the instability of the mode; it is
absent for the case of a ghost with a stable dispersion relation 
when the pole
lies at positive $\omega^2$. Thus we conclude that for a mode with
gradient instability there is no unambiguous way to tell if it is a
ghost or not. Rather, this notion makes sense only for a given
coupling of the mode to the source.

It is instructive to trace explicitly the agreement of one-khronon
exchange amplitudes calculated in the St\"uckelberg and unitary gauge
pictures. 
To this aim we need to specify the source. This is easier to do
on the 
St\"uckelberg side where we couple the
khronon to a scalar field $\Sigma$. The khronon field must enter
with derivatives so we write,
\be
\label{khronosource}
S_{source}=\int d^4x \sqrt{-g} g^{\mu\nu}\d_\mu\varphi\d_\nu\Sigma\;.
\ee 
In terms of the khronon perturbations this takes the form 
\[
S_{source}=\int d^4x\; (-\ddot\chi+\Delta\chi)\,\Sigma\;.
\]
From this expression and the quadratic action (\ref{SchiIII})
one reads off the khronon exchange amplitude
\be
\label{sigmaamp}
{\cal A}_{\Sigma}\propto\Sigma\;\frac{(\omega^2-{\bf p}^2)^2}{D}\;\Sigma\;,
\ee
where 
\[
%\label{D}
D=M^2_{\lambda_1}\omega^4+M^2_\alpha\omega^2{\bf p}^2
-M_\lambda^2{\bf p}^4+2M^4\omega^2\;.
\]
One observes that this amplitude is a sum of two contributions,
with positive and negative residues at the poles. In this sense one of
the khronon modes is indeed a ghost.

Let us now see how the same amplitude is recovered in the unitary
gauge. Fixing $\varphi=t$ and
substituting the decomposition (\ref{dec}) into the source term
(\ref{khronosource}) we obtain 
\[
S_{source}=\int d^4x\;(-\dot\phi+2\dot\psi+\dot E+\sqrt\Delta B)
\,\Sigma\;.
\]
The next step is to integrate out the non-dynamical fields $B$,
$E$. Importantly, this produces a
contribution into the Lagrangian which is quadratic in
$\Sigma$. Omitting the higher-derivative terms the
resulting Lagrangian reads 
\[
{\cal L}^{(2)}_{III}=\frac{M^2_\lambda}{2}\dot{\tilde\psi}^2
+M^2_\lambda\phi\Delta\tilde\psi+\frac{M_\alpha^2}{2}(\d_i\phi)^2
+\frac{M^2_{\lambda_1}}{2}\dot\phi^2+M^4\phi^2
-\big(\dot\phi+\dot{\tilde\psi}\big)\Sigma
+\frac{\Sigma^2}{2M_\lambda^2}\;,
\]
where for simplicity we have taken the limit $M_P\to\infty$ (this
corresponds to the decoupling limit in the St\"uckelberg picture) 
and introduced the rescaled variable
\[
\tilde\psi=(2M_P^2/M_\lambda^2)\;\psi\;.
\]
This Lagrangian leads to the following
propagators:
%\bseq 
\begin{align}
&\langle \phi\,\phi\rangle=\frac{\omega^2}{D}\;,~~~~~~~
\langle \phi\,\tilde{\psi}\rangle=-\frac{{\bf p}^2}{D}\;,\notag\\
&\langle \tilde\psi\,\tilde\psi\rangle=
\frac{M^2_{\lambda_1}\omega^2+M^2_\alpha{\bf p}^2+2M^4}
{M^2_\lambda\, D}\;.\notag
\end{align} 
%\eseq
Finally, the khronon exchange amplitude reads,
\[
{\cal A}_\Sigma\propto\Sigma\;\big(
\omega^2(\langle \phi\,\phi\rangle+2\langle \phi\,\tilde{\psi}\rangle
+\langle \tilde\psi\,\tilde\psi\rangle)-M_\lambda^{-2}\big)\;\Sigma\;.
\]
Combining everything together one obtains the result
(\ref{sigmaamp}). Thus we find that the unitary gauge calculation
leads (in a quite non-trivial way) to the same result as in the
St\"uckelberg formalism, as, of course, it should be. Technically,
in the unitary gauge the ghost pole appears due to the
structure of the coupling to the source.

\section{Stability bounds for the healthy model}
\label{app:C}

Here we present conditions on the parameters of the
healthy model of Sec.~\ref{sec:healthy}
imposed by requiring the stability of the scalar mode.
We formulate them in terms of the coefficients $\alpha$, $f_n$, $g_n$
appearing in the quadratic Lagrangian (\ref{L2imp}). 

The stability
requirement is expressed by Eq.~\eqref{condPQ} with the polynomials
$P(x)$ and $Q(x)$ defined in \eqref{poly1},
\eqref{poly2}.
This implies the following necessary conditions:
\be
\label{eq:nec}
\begin{split}
&  g_2^2-g_1 g_3>0\;,~~~~ 2-\alpha >0\;,\\
& g_3 >0\;,~~~~~  \alpha >0\;,~~~~~ f_3 < 2 \sqrt{\alpha\, g_3}.
\end{split}
\ee
Deriving the full set of necessary and sufficient 
conditions is quite cumbersome. Instead we 
provide two different sets of sufficient conditions. The first
possibility is requiring
  that all the monomials in $P(x)$ are positive definite.
  Apart from the previous conditions \eqref{eq:nec} this yields the
  constraints
\be
\begin{split}
\label{eq:suf1}
&  g_1 f_3 + g_3 f_1 - 2 g_2 f_2 >0\;,\\
&  f_2^2 - 4 g_2 - f_1 f_3 -2 g_3 -\alpha g_1 >0\;,\\
&  2 f_3 + \alpha f_1 + 4 f_2 >0\;,
\end{split}
\ee

Another option is to write $P(x)$ as
\[
%\label{eq:quartic}
\begin{split}
&P(x)=\left((g_2^2-g_1
g_3)^{1/4}x+(4-2\alpha)^{1/4}\right)^4-x\Big(c_2x^2-c_1x+c_0\Big),
\end{split}
\]
and require the quadratic polynomial inside the last bracket 
to be positive at $x<0$. This translates
into the constraints 
\be
\label{eq:suf2}
\begin{split}
c_2>0\;, \ \ c_1>-2\sqrt{c_0 c_2}\;,\ \ c_0>0\;, 
\end{split}
\ee
where
\[
\begin{split}
&c_2=f_3 g_1+f_1 g_3-2 f_2 g_2+4(g_2^2-g_1g_3)^{3/4}(4-2\alpha)^{1/4},\\
&c_1=f_2^2-4 g_2-f_1 f_3-2g_3-g_1\alpha-6
(g_2^2-g_1g_3)^{1/2}(4-2\alpha)^{1/2},\\
&c_0=2f_3+4f_2+f_1\alpha
+4 (g_2^2-g_1g_3)^{1/4}(4-2\alpha)^{3/4}\;.
\end{split}
\]
Note that the two sets of bounds (\ref{eq:suf1}) and (\ref{eq:suf2})
overlap but none of them contains the other.

Clearly, the necessary conditions (\ref{eq:nec}) can be complemented
with any of the sufficient
conditions (\ref{eq:suf1}) and (\ref{eq:suf2}).
To demonstrate that the parameter space restricted by the stability
bounds is not empty let us give an explicit example. It is
straightforward to verify that the set of parameters
\[
\alpha=g_3=1.5\,,~~~~~
f_1= f_2=-f_3=-g_1=-g_2=2
\]
satisfies the constraints (\ref{eq:nec}), (\ref{eq:suf1}) and thus
leads to stable dispersion relation of the scalar mode.

The bounds presented above can be
translated directly in terms of the parameters $A_i$, $B_i$, $C_i$ and $D_i$
 in the original potential 
\eqref{potadd} but we do not do it here.

\section{Spherically symmetric solutions in  Einstein-aether and
khrono-metric theories}
\label{app:F}

In this appendix we demonstrate that 
spherically symmetric solutions 
of the khrono-metric model (\ref{covar22}) are identical to those of 
the Einstein-aether theory.
Let us consider the
equation of motion for the
khronon field coming from varying the action
(\ref{covar22}) with respect to the field $\chi$
\be
\label{Jcons}
\nabla_\mu J^\mu=0\;,
\ee
where
\[
%\label{Jmu}
J^\mu=\frac{P^\mu_\nu}{\sqrt X}\;\frac{1}{\sqrt{-g}}\frac{\delta
   S}{\delta u_\nu}\;.
\]
At the same time the equation of motion for the aether is obtained by varying
(\ref{covar22}) with respect to the field $u_\n$ and reads
\be
\label{aethereq}
P^\mu_\nu\frac{1}{\sqrt{-g}}\frac{\delta
   S}{\delta u_\nu}=0\;.
\ee
In deriving this equation one has to take into account
the constraint $u_\mu u^\mu=1$: it leads to the appearance of the
   projector $P^\mu_\nu$ on the l.h.s. 
Finally, the energy-momentum tensor appearing in the Einstein
equations
for the khrono-metric theory can
be written
as
\[
T_{\m\n}=\frac{2}{\sqrt{-g}}\left(\frac{\delta S}{\delta
g^{\m\n}}\bigg|_{u_\s}-\frac{1}{2}\frac{\delta S}{\delta u_{\s}}
u^\sigma u_\mu u_\nu\right)\;,
\]
where the second term comes from
the explicit dependence of the
vector $u_\mu$ in the khronon theory on the metric, see
Eqs.(\ref{umu}), (\ref{X}). This
coincides with the energy-momentum tensor of the aether. To obtain the
second term in this case one again has to take into account the
constraint $u_\mu u^\mu=1$ \cite{Jacobson:2010mx}.

Any spherically
symmetric configuration of aether is automatically
hy\-per\-sur\-face-or\-tho\-go\-nal 
implying that any spherically symmetric solution
of the Einstein-aether theory is a solution for the khrono-metric theories
\cite{Jacobson:2010mx}. The converse is less obvious as the equation
of motion (\ref{Jcons}) of the khronon field contains 
an additional derivative compared to the aether 
equation (\ref{aethereq})
and
thus, a priori, admits more solutions. However,
  for spherically symmetric configurations
(\ref{Jcons}) implies (\ref{aethereq}).
Indeed, the current $J^\mu$ obeys the relation
\[
u_\mu J^\mu=0\;.
\]
Hence in the unitary gauge (\ref{unitg}) its component $J^0$
identically
vanishes imlying that
the corresponding charge
\[
Q\equiv\int \di^3x\sqrt\gamma J^0\;
\]
is identically zero. On the other hand,
the time derivative of $Q$ is equal to the
flux of the spatial component $J^i$ of the current through the
2-sphere at spatial infinity\footnote{We assume that the 3d surfaces
   $\varphi=const$ do not
   have holes. This is the case if these surfaces form a
   regular foliation of the whole space-time.}.
In spherically symmetric situation this implies that the
current itself is zero which brings us to the equation
(\ref{aethereq}). Combining this result with 
the equality of the energy-momentum tensors 
we conclude that spherically symmetric
solutions in the khronon and Einstein-aether theories are indeed
identical.

\section{PPN parameters $\alpha_1^{PPN}$,
  $\alpha_2^{PPN}$ for the healthy model} 
\label{app:D}

In this appendix we derive the formulas for the PPN 
parameters $\alpha_1^{PPN}$,
$\alpha_2^{PPN}$ in the khronon theory (\ref{covar22}).
We assume the couplings $\alpha,\beta,\lambda'$ to be small and
perform calculations to the leading order in these couplings.
We consider the metric produced by a
point source of mass $m$ {\em in its rest frame}. This frame does not
coincide with the frame defined by the preferred foliation. Hence the
background value $\bar\varphi$ of the khronon field in this frame
differs from 
the coordinate time. Using the reparameterization symmetry
(\ref{phirepar}) we fix 
\[
%\label{barphi}
\bar\varphi=\sqrt{1+v^2}\,t+v^ix^i\;,
\]
where $v^i$ is the velocity of the source with respect to the
preferred frame. This corresponds to the background value of the vector
$u_\mu$
\[
%\label{ubar}
\bar u_0=\sqrt{1+v^2}~,~~~\bar u_i=v^i\;.
\]
The source perturbs the metric and the khronon. One writes,
\[
%\label{pert}
g_{\mu\nu}=\eta_{\mu\nu}+h_{\mu\nu}~,~~~\varphi=\bar\varphi+\chi\;.
\]
To the leading order $h_{\mu\nu}$ is given by the standard 
Newtonian expressions,
\be
\label{hN}
h_{00}^{(0)}=2\phi(r)~,~~~h_{0i}^{(0)}=0
~,~~~h_{ij}^{(0)}=2\phi(r)\delta_{ij}\;,
\ee
where
\be
\label{phiN}
\phi(r)=-\frac{m}{8\pi M_P'^2 r}\;.
\ee
Our goal is to find corrections to (\ref{hN}) in powers of $v$.

It is convenient to introduce the following notations
\begin{gather}
\label{Phi}
\Phi=\frac{1}{2}\bar u^\mu\bar u^\nu h_{\mu\nu}~,~~~~~
V_\rho=\bar u^\mu\bar P^\nu_\rho h_{\mu\nu}~,~~~~~
H_{\lambda\rho}=\bar P^\mu_\lambda\bar P^\nu_\rho h_{\mu\nu}\;,\\
%\label{dpar}
\dpa=\bar u^\mu\d_\mu~,~~~~~
\dpe_\mu=\bar P^\nu_\mu\d_\nu~,~~~~~
\Bpe=\dpe_\nu{\dpe}^\nu\;.\notag
\end{gather}
The indices here are raised and lowered using the Minkowski metric
$\eta_{\mu\nu}$. 
Expanding to linear order in perturbations we obtain
\begin{align}
%\label{ulin}
&u_\mu=\bar u_\mu+\dpe_\mu\chi+\bar u_\mu\Phi\;,\notag\\
%\label{dulin}
&\nabla_\nu u_\mu=\bar u_\nu\dpa\dpe_\mu\chi
+\dpe_\nu\dpe_\mu\chi-\bar u_\nu\dpe_\mu\Phi
-\frac{1}{2}\dpe_\nu V_\mu
-\frac{1}{2}\dpe_\mu V_\nu
+\frac{1}{2}\dpa H_{\mu\nu}\;.\notag
\end{align}
Substituting these expressions into the action for the khronon sector
(last three terms in (\ref{covar22}))
we obtain at the quadratic level,
\be
\label{Schilin}
\begin{split}
S_\chi=-\frac{M_P'^2}{2}\int \di^4 x\bigg\{
\beta\bigg(&(\Bpe\chi)^2-2\Bpe\chi\dpe_\nu V^\nu
-\dpa\chi\dpe_\nu\dpe_\mu H_{\mu\nu}
+\frac{1}{2} (\dpe_\nu V^\nu)^2\\
&+\frac{1}{2} \dpe_\nu V_\mu {\dpe}^\nu V^\mu
-\dpa V_\mu \dpe_\nu H^{\mu\nu}
+\frac{1}{4}\dpa H_{\mu\nu}\dpa H^{\mu\nu}\bigg)\\
&+\lambda'\bigg(\Bpe\chi-\dpe_\nu V^\nu+\frac{1}{2}\dpa H\bigg)^2
+\alpha\Big(\dpa\dpe_\mu\chi-\dpe_\mu\Phi\Big)^2\bigg\}\;,
\end{split}
\ee
where $H=H_\nu^\nu$\;. This yields the equation for
the khronon perturbation $\chi$,
\be
\label{chieqlin}
(\lambda'+\beta)(\Bpe)^2\chi+\alpha(\dpa)^2\Bpe\chi=
\alpha\dpa\Bpe\Phi+(\lambda'+\beta)\Bpe\dpe_\nu V^\nu
-\frac{\lambda'}{2}\dpa\Bpe H
-\frac{\beta}{2}\dpa\dpe_\mu\dpe_\nu H^{\mu\nu}.
\ee
Variation of the action (\ref{Schilin}) with respect to the metric
perturbation $h_{\mu\nu}$ gives linearized khronon 
energy-momentum tensor:
\be
\label{Tchilin}
\begin{split}
T_\chi^{\mu\nu}&=-2\frac{\delta S_\chi}{\delta h_{\mu\nu}}\\
&=-\bar u^\mu\bar u^\nu\frac{\delta S_\chi}{\delta\Phi}
-(\bar u^\mu\bar P^\nu_\lambda+\bar u^\nu\bar P^\mu_\lambda)
\frac{\delta S_\chi}{\delta V_\lambda}
-2\bar P^\mu_\lambda\bar P^\nu_\rho
\frac{\delta S_\chi}{\delta H_{\lambda\rho}}\;,
\end{split}
\ee
where we have used the decomposition
\[
%\label{hPhi}
h_{\mu\nu}=2\bar u_\mu\bar u_\nu\Phi+\bar u_\mu V_\nu
+\bar u_\nu V_\mu+H_{\mu\nu}\;.
\]
Evaluating the variations entering into (\ref{Tchilin}) we obtain
\bseq
\label{dS}
\begin{align}
\label{dSPhi}
\frac{\delta S_\chi}{\delta\Phi}=-M_P'^2\alpha&
\Big(\dpa\Bpe\chi-\Bpe\Phi\Big)\;,\\
\frac{\delta S_\chi}{\delta V_\lambda}=-M_P'^2\bigg[&
(\lambda'+\beta){\dpe}^\lambda\Bpe\chi
-\left(\lambda'+\frac{\beta}{2}\right){\dpe}^\lambda\dpe_\rho V^\rho \notag\\
&-\frac{\beta}{2}\Bpe V^\lambda+\frac{\lambda'}{2}{\dpe}^\lambda\dpa H
+\frac{\beta}{2}\dpa\dpe_\rho H^{\lambda\rho}\bigg]\;,
\label{dSV}\\
\frac{\delta S_\chi}{\delta H_{\lambda\rho}}=-M_P'^2\bigg[&
-\frac{\lambda'}{2}\eta^{\lambda\rho}\bigg(\dpa\Bpe\chi
-\dpa\dpe_\sigma V^\sigma+\frac{1}{2}(\dpa)^2H\bigg)\notag\\
&-\frac{\beta}{2}\dpa{\dpe}^\lambda{\dpe}^\rho\chi
+\frac{\beta}{4}\dpa{\dpe}^\lambda V^\rho
+\frac{\beta}{4}\dpa{\dpe}^\rho V^\lambda
-\frac{\beta}{4}(\dpa)^2 H^{\lambda\rho}\bigg]\;.
\label{dSH}
\end{align}
\eseq

The rest of the calculation proceeds as follows.
One inserts the Newtonian metrics (\ref{hN}) into Eqs.~(\ref{Phi}) and
find the potentials  
$\Phi^{(0)}$, $V_\rho^{(0)}$ and
$H_{\lambda\rho}^{(0)}$. The latter act as a source for the
khronon perturbation $\chi$ in Eq.~(\ref{chieqlin}). 
At the next step one combines the khronon perturbation found from
(\ref{chieqlin}) and the Newtonian expressions for the potentials
into the khronon energy-momentum tensor (\ref{Tchilin}). This tensor 
substituted in the r.h.s. of the Einstein's equations
determines the correction to the metric:
\be
\label{Einslin}
\Delta h^{(1)}_{\mu\nu}=\frac{2}{M_P'^2}\left(T_{\chi\;\mu\nu}-
\frac{1}{2}\eta_{\mu\nu}T_\chi\right)\;.
\ee
Here we have imposed the harmonic gauge,
\[
\d_\mu h^{\mu\nu}-\frac{1}{2}\d^\nu h=0
\]
and have used the fact that the metric is static. Note that the first
order in the post-Newtonian approximation requires to find
$h^{00}$, $h^{0i}$ and $h^{ij}$ components of the metric with the
accuracy $O(v^2)$, $O(v)$ and $O(1)$ respectively. This
implies that we need to determine 
$T^{\mu\nu}_\chi$ and $T_\chi$ to order $O(v^2)$, 
$T_\chi^{0i}$ -- to order $O(v)$, and $T^{ij}_\chi$ -- to order
$O(1)$.

Expanding up to terms $O(v^2)$ we obtain:
%\bseq
%\label{uP}
\begin{gather}
\bar u_0=\bar u^0=1+v^2/2~,~~~\bar u_i=-\bar u^i=v^i\;,\notag\\
\bar P^0_0=-v^2~,~~~\bar P^i_0=-\bar P^0_i=v^i~,~~~\bar
P^i_j=\delta^i_j+v^iv^j\;,\notag\\
\dpa=-v^i\d_i~,~~~\dpe_0=v^i\d_i~,~~~\dpe_i=\d_i+v^iv^j\d_j\;\notag\\
\Bpe=-\Delta-v^iv^j\d_i\d_j\;,\notag
\end{gather}
%\eseq
where we have used that derivatives act on static
configurations. Substituting these expressions together with
(\ref{hN})
into (\ref{Phi}) 
we find
%\bseq
%\label{PhiH}
\begin{gather}
\Phi^{(0)}=(1+2v^2)\phi(r)\;,\notag\\
V_0^{(0)}=-4v^2\phi(r)~,~~~V_i^{(0)}=-4v^i\phi(r)\;,\notag\\
H_{00}^{(0)}=2v^2\phi(r)~,~~~H_{0i}^{(0)}=2v^i\phi(r)~,~~~
H_{ij}^{(0)}=(2\delta_{ij}+6v^iv^j)\phi(r)\;,\notag\\
H^{(0)}=(-6-4v^2)\phi(r)\notag
\end{gather}
%\eseq
The khronon equation (\ref{chieqlin}) takes the form,
\be
\label{chieqlin1}
(\lambda'+\beta)\Delta^2\chi=(\alpha-\lambda'-3\beta)v^i\d_i\Delta\phi(r)\;,
\ee
where on the l.h.s we have neglected terms
$O(v^2\Delta^2\chi)$ as they are of higher order in
$v$. From (\ref{chieqlin1}) we find
\[
%\label{chilin}
\Delta\chi=\frac{\alpha-\lambda'-3\beta}{\lambda'+\beta}v^i\d_i\phi(r)\;,
\]
meaning that $\chi$ is of order $O(v)$. 

Let us estimate the orders of the
variations (\ref{dS}). One finds by inspection that 
$\frac{\delta S_\chi}{\delta\Phi}$ is of order $O(1)$, 
$\frac{\delta S_\chi}{\delta V_\lambda}$ -- at most of order $O(v)$, 
$\frac{\delta S_\chi}{\delta H_{\lambda\rho}}$ -- at most of order
$O(v^2)$. This allows to simplify the
khronon energy-momentum tensor. To the required orders we have:
\bseq
\label{Ts}
\begin{align}
&T_{\chi\;00}=-(1+v^2)\frac{\delta S_\chi}{\delta\Phi}
+2v^i\frac{\delta S_\chi}{\delta V_i}\;,\\
&T_\chi=-\frac{\delta S_\chi}{\delta\Phi}+
2\delta_{ij}\frac{\delta S_\chi}{\delta H_{ij}}\;,\\
&T_{\chi\;0i}=-v^i\frac{\delta S_\chi}{\delta\Phi}+
\frac{\delta S_\chi}{\delta V_i}\;,\\
&T_{\chi\;ij}=0\;.
\end{align}
\eseq
One evaluates the variations appearing in these formulas,
\bseq
\label{vars}
\begin{align}
&\frac{\delta S_\chi}{\delta\Phi}=-M_P'^2\alpha
\bigg[(1+2v^2)\Delta\phi(r)+\frac{\alpha-2\beta}{\lambda'+\beta}
v^iv^j\d_i\d_j\phi(r)\bigg]\;,\\
&\frac{\delta S_\chi}{\delta V_i}=-M_P'^2\Big[
2\beta v^i\Delta\phi(r)+(\alpha-2\beta)v^j\d_i\d_j\phi(r)\Big]\;,\\
&\delta_{ij}\frac{\delta S_\chi}{\delta H_{ij}}=-M_P'^2
\frac{(\alpha-2\beta)(3\lambda'+\beta)}{2(\lambda'+\beta)}
v^iv^j\d_i\d_j\phi(r)\;.
\end{align}
\eseq
Inserting (\ref{vars}) into (\ref{Ts}) and substituting the result
into (\ref{Einslin}) we find the equations for the
first order corrections to the metric,
\bseq
\begin{align}
\label{h00PN}
&\Delta h_{00}^{(1)}=\big(\alpha+4(\alpha-2\beta)v^2\big)\Delta\phi(r)+
\frac{(\alpha-2\beta)(\alpha-\lambda'-3\beta)}{\lambda'+\beta}
v^iv^j\d_i\d_j\phi(r)\;,\\
\label{h0iPN}
&\Delta h_{0i}^{(1)}=2(\alpha-2\beta)v^i\Delta\phi(r)
-2(\alpha-2\beta) v^j\d_i\d_j\phi(r)\;,\\
\label{hijPN}
&\Delta h_{ij}^{(1)}=\alpha\delta_{ij}\Delta \phi(r)\;.
\end{align}
\eseq
It is straightforward to solve these equations for the explicit form
(\ref{phiN}) of the function $\phi(r)$. Note that the second term on
the r.h.s. of (\ref{h0iPN}) can be removed by a time-independent
gauge transformation. Combining the result with the Newtonian
expressions (\ref{hN}) we obtain the metric (\ref{hPPN})
with 
\be
\label{GN1}
G_N=\frac{1}{8\pi M_P'^2}\left(1+\frac{\alpha}{2}\right)
\ee
and the PPN parameters $\alpha_1^{PPN}$, $\alpha_2^{PPN}$ quoted in
(\ref{PPNpar}). The expression (\ref{GN1}) coincides with
(\ref{GN}) to linear order in $\alpha$.

\end{document}